\documentclass{aastex6}

\usepackage{graphicx}
\usepackage{epstopdf}
\usepackage{amsmath}

\newcommand{\lya}{\mbox{Lyman-$\alpha$}}
\newcommand{\ovi}{{\sc O~vi}}
\newcommand{\hi}{{\sc H~i}}
\newcommand{\rsun}{$R_\odot$}
\newcommand{\kms}{km~s$^{-1}$}
\newcommand{\avg}[1]{\left\langle #1 \right\rangle}

\begin{document}
\title{Determination of Coronal Mass Ejection physical parameters from combination of polarized visible light and UV Lyman-$\alpha$ observations}
\author{R. Susino \and A. Bemporad}
\affil{Istituto Nazionale di Astrofisica (INAF), Osservatorio Astrofisico di Torino, Via Osservatorio 20, I-10025 Pino Torinese, Torino, Italy}
\shortauthors{Susino \& Bemporad}
\shorttitle{Determination of CME physical parameters from visible-light and UV observations}

\begin{abstract}
Visible-light observations of Coronal Mass Ejections (CMEs) performed with coronagraphs and heliospheric imagers (in primis on board the SOHO and STEREO missions) have offered so far the best way to study the kinematics and geometrical structure of these fundamental events. 
Nevertheless, it has been widely demonstrated that only combination of multi-wavelength data (including X-ray spectra, EUV images, EUV-UV spectra, and radio dynamic spectra) can provide complete information on the plasma temperature and density distributions, non-thermal motions, magnetic fields, and other physical parameters, for both CMEs and CME-related phenomena.
In this work, we analyze three CMEs by combining simultaneous data acquired in the polarized visible light by the LASCO-C2 coronagraph and in the UV \hi\ \lya\ line (1216~\AA) by the UVCS spectrometer, in order to estimate the CME plasma electron density (using the polarization-ratio technique to infer the 3D structure of the CME) and temperature (from the comparison between the expected and measured \lya\ intensities) along the UVCS field of view.
This analysis is primarily aimed at testing the diagnostic methods that will be applied to coronagraphic observations of CMEs delivered by the Metis instrument on board the next ESA-Solar Orbiter mission.
We find that CME cores are usually associated with cooler plasma ($T \sim 10^6$~K), and that a significant increase of the electron temperatures is observed from the core to the front of the CME (where $T > 10^{6.3}$~K), which seems to be correlated, in all cases, with the morphological structure of the CME as derived from visible-light images.
\end{abstract}

\maketitle
\section{Introduction}
During major solar eruptions (or Coronal Mass Ejections -- CMEs), huge bubbles of highly ionized plasma (often associated with shock waves and Solar Energetic Particles -- SEPs) expand into the interplanetary space, affecting a significant fraction of the whole heliosphere, and eventually propagating even out of the heliopause as interstellar shocks in the most dramatic cases, as recently reported with Voyager~1 observations \citep{gurnett2015}. 
Solar transients (flares, CMEs and prominence eruptions) have an impact on all the planetary objects, interacting with their magnetospheres or magnetospheric-like structures and inducing geomagnetic storms \citep[see review by][]{akasofu_2011}, as well as beautiful auroras on Earth and other planets \citep{hultqwist_2008}. 
These events likely played a role even in the development of life on Earth, for instance by modulating the rate of galactic cosmic rays impacting on the early Earth's atmosphere \citep[via the well-known ``Forbush decrease'' effect; see][]{lockwood1071} and on the atmospheric chemistry \citep{airapetian2016}. 
Moreover, the study of these events is very important also from a theoretical point of view, because in order to understand their origin and interplanetary evolution it is necessary to consider many different plasma physical processes (such as plasma instabilities, magnetic reconnections, wave-particle interactions, etc.), phenomena that are only partially understood.

From the observational point of view, the study of solar eruptions can be performed using many different data delivered daily by both ground- and space-based observatories. 
Nevertheless, images of the solar disk can provide only information on the location of the source region, on the eruption start time and on the early expansion phases. 
Then, after the eruptions take off and leave the Sun, their subsequent evolution during the expansion and propagation phases can be followed only with two classes of instruments: space-based coronagraphs and heliospheric imagers. 
In fact, these are the only instruments covering the huge amount of space traveled by solar eruptions during their propagation from the Sun to the Earth and beyond. 
Without data provided by space-based coronagraphs and heliospheric imagers it would be simply impossible to characterize the real CME propagation angle and CME speed, and to investigate the physical processes occurring during their interplanetary expansion.

It is well known today that, after the main impulsive acceleration phase occurring in the lower corona, solar eruptions are subject to many different processes affecting their evolution, which is never like a simple radial and self-similar expansion. 
During their early propagation phases, CMEs are often channelled by coronal streamers and/or deflected away from nearby coronal holes \citep[see][and refeences therein]{mostl2015}, or towards the interplanetary current sheet \citep[e.g.,][]{byrne2010,isavnin2014}; these interactions may modify their propagation directions up to 25$^\circ$ with respect to the location of the source region \citep{kay2013}, influencing the strength of the eventual impact on Earth. 
Significant rotations of CMEs around their propagation axis are also observed in many cases \citep[e.g.,][]{thompson2012,bemporad2011}, which change the orientation of the magnetic field of the associated magnetic cloud impacting on the Earth's magnetosphere, and in turn their capability to induce geomagnetic storms. 
Moreover, significant magnetic drag occurs during the interplanetary propagation of CMEs, leading to further accelerations or decelerations of the ejecta (depending on the expansion speed relative to the ambient solar wind) that affect the expected arrival times at Earth \citep[e.g.][]{iju2014,temmer2011}. 
Furthermore, when multiple eruptive events are ejected in sequence, CME-CME interactions may occur increasing their final geoeffectiveness \citep[see][]{far06,wu07}.

All these phenomena make CME observations by coronagraphs and heliospheric imagers crucial for understanding these events and forecasting their impact on the Earth. 
Moreover, previous experience with visible-light (VL) coronagraphs shows that unique information can be derived only when data acquired at different wavelengths are combined together, i.e., not only images in the VL, but also X-ray spectra, EUV images, EUV-UV spectra, and radio dynamic spectra. 
In particular, the combination of observations acquired in the VL by different coronagraphs and in the UV spectral range by the UVCS spectrometer \citep{kohl1995} on SOHO, allowed to characterize the distribution of plasma temperatures (electron and ion) and their evolution inside the CME core and front, to study many CME-related phenomena, such as post-CME current sheets and CME-driven shocks, and to reveal the three-dimensional (3D) CME structure \citep[see][for an extensive review of these results]{koh06}. 
More recently, UVCS spectra have been combined for the first time with SOHO and STEREO VL images to perform the first stereoscopic and spectroscopic reconstruction of a CME \citep{susino14}, to derive the physical parameters of coronal plasma -- including the magnetic field -- across CME-driven shocks \citep[e.g.,][]{bemporad14,susino15}, and to derive kinetic temperature, gas pressure, and filling factor in erupting prominences \citep{heinzel16}. All these works demonstrate the importance of complementarity of VL and UV observations of solar eruptions.

In the near future, combined VL and UV images will be provided by the Metis coronagraph \citep{antonucci12,fineschi12,romoli14} on board the ESA-Solar Orbiter mission, due to launch in October 2018. 
The Metis instrument will acquire the first-ever simultaneous observations of the solar corona in the polarized visible light (broad-band 580-640~nm) and in the UV (narrow-band around the \hi\ \lya\ 1216~\AA\ line), even if, since 2012, the instrument has lost its spectroscopic capabilities. 
Hence, the aim of the present work is to perform the first tests on the diagnostic capabilities for CMEs that will be possible with future Metis data. 
This test is performed here using available simultaneous observations of real CMEs acquired in the polarized visible light by the SOHO/LASCO-C2 coronagraph \citep{brueckner95} and in the UV by the UVCS spectrometer. 
Because the Metis coronagraph will not provide spectroscopic observations, in this work we focused only on the observed evolution of UV \lya\ intensities, thus simulating the information that will be provided by Metis. 
The paper is organized as follows: after a description of the selected events and datasets (Section 2), we describe the diagnostics (Section 3) we applied for the determination of CME electron density (Section 3.1) and electron temperature (Section 3.2), and then we discuss our results (Sections 4 and 5).

\section{Observations}
For the purpose of this work we looked for CME events for which simultaneous measurements of visible-light polarized brightness ($pB$) by LASCO-C2 and \hi\ Lyman-$\alpha$ spectra by UVCS were available.

\begin{figure*}[t]
\caption{\label{fig_events}{\em Right column:} SOHO/EIT 304~\AA\ images acquired before the CMEs occurred on 2000 November 8 (top row), 2000 December 25 (middle row), and 2003 May 2-3 (bottom row). {\em Middle and left columns:} LASCO-C2 total brightness images (processed with the NRGF in order to better highlight the CME structure) at the time of the $pB$ measurement and several hours later, respectively, showing the evolution of the three CMEs in the coronagraph field of view. In the central column, the SOHO/UVCS field of view (the slit, represented as a red line) is also superimposed to mark the portion of the solar corona sampled by the spectrometer during the transit of the CME.}
\centering
\includegraphics[width=\textwidth]{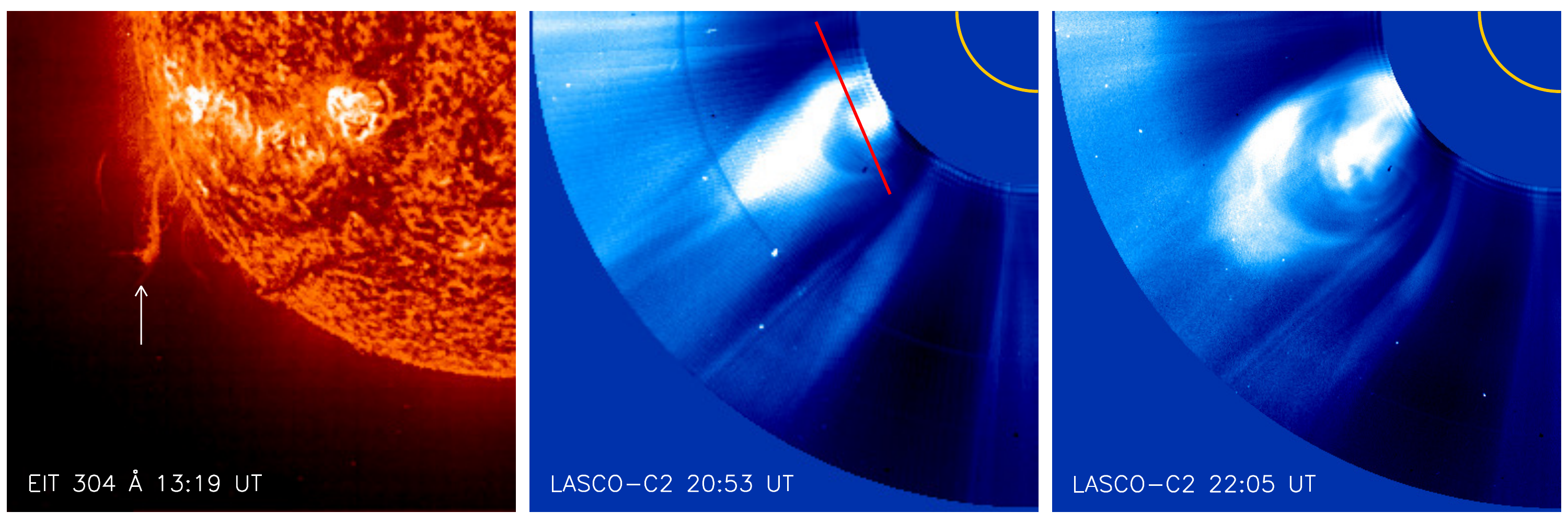}
\includegraphics[width=\textwidth]{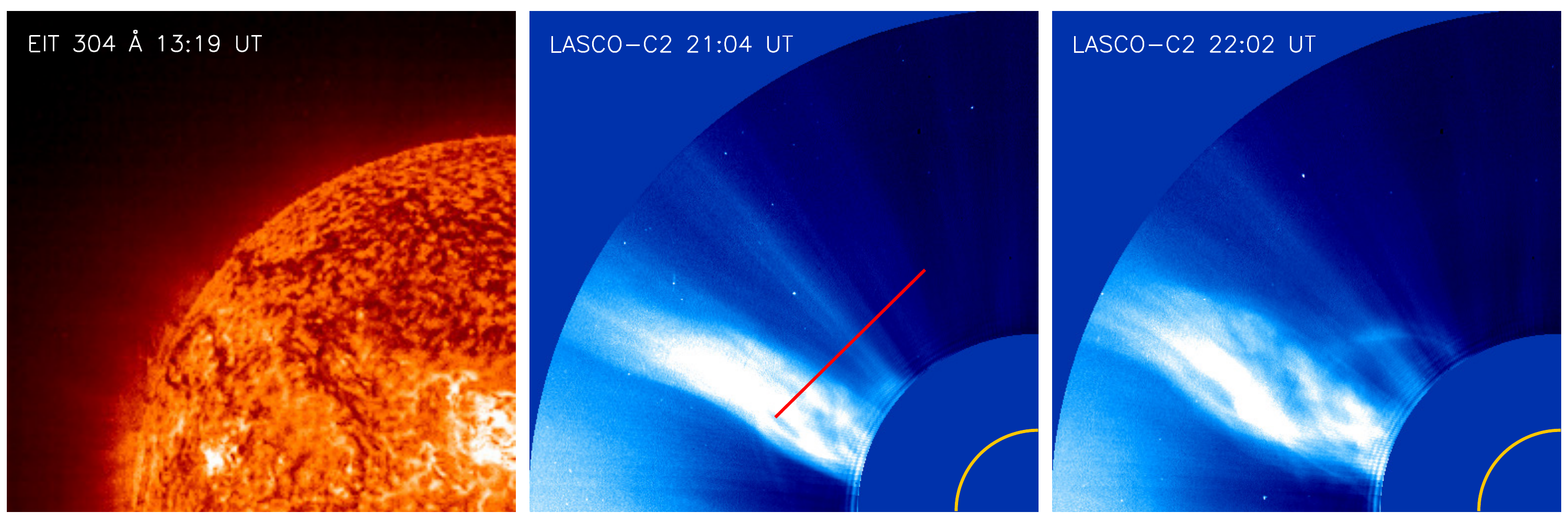}
\includegraphics[width=\textwidth]{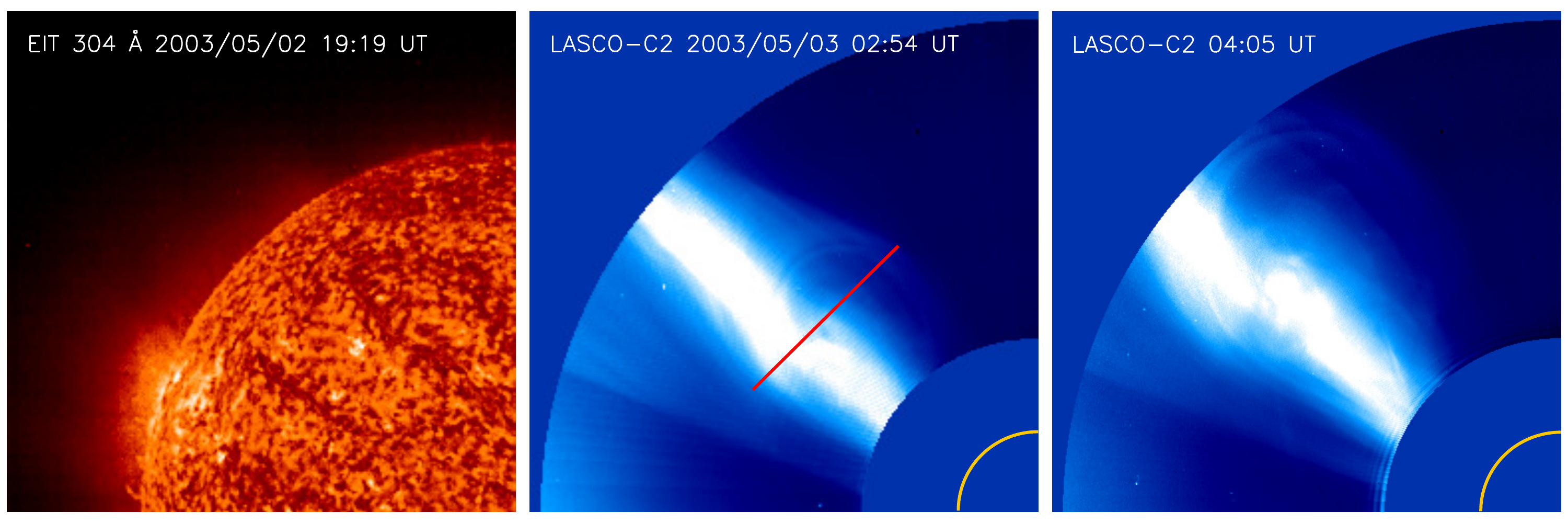}
\end{figure*}

In order to find suitable events, we first searched the LASCO CME catalog for all the CMEs observed with LASCO-C2 during the acquisition of a polarimetric sequence for the measurements of the $pB$; then we made a cross-search of the UVCS observation logs to retrieve, among those, the events that were simultaneously observed by UVCS, selecting only those where significant part of the CME structure crossed the UVCS field of view (FOV).
The probability to have good candidates is quite low, mostly because, on one hand LASCO-C2 acquires only one or two polarimetric sequences per day, i.e., with a very poor cadence, on the other hand, UVCS detectors started to be progressively more and more degraded since year 2006, with reduced FOV. 
Moreover, for a large amount of time UVCS observations were performed with the slit centered at projected altitudes below the inner edge of the LASCO-C2 occulter ($\sim 2.2$ solar radii), thus preventing the possibility to analyze erupting events at the same location and at the same time with the two instruments. 
As a result, only three events in the whole SOHO era were found. All the three selected CMEs consist in minor events with small plasma eruptions at quite low velocities, as we will describe in the following.

\subsection{Description of the events}\label{events}

\begin{figure*}[t]
\caption{\label{event_1}Mauna Loa Mark-{\sc iv} images of the VL corona (processed with the NRGF) showing the initial evolution of Event~1. The red, dashed line marks the inner edge of the LASCO-C2 field of view (at 2.2~\rsun). Note the clear three-part structure of the CME with the bright core, the dark void, and the almost hemispherical front surrounding it.}
\centering
\includegraphics[width=0.66\textwidth]{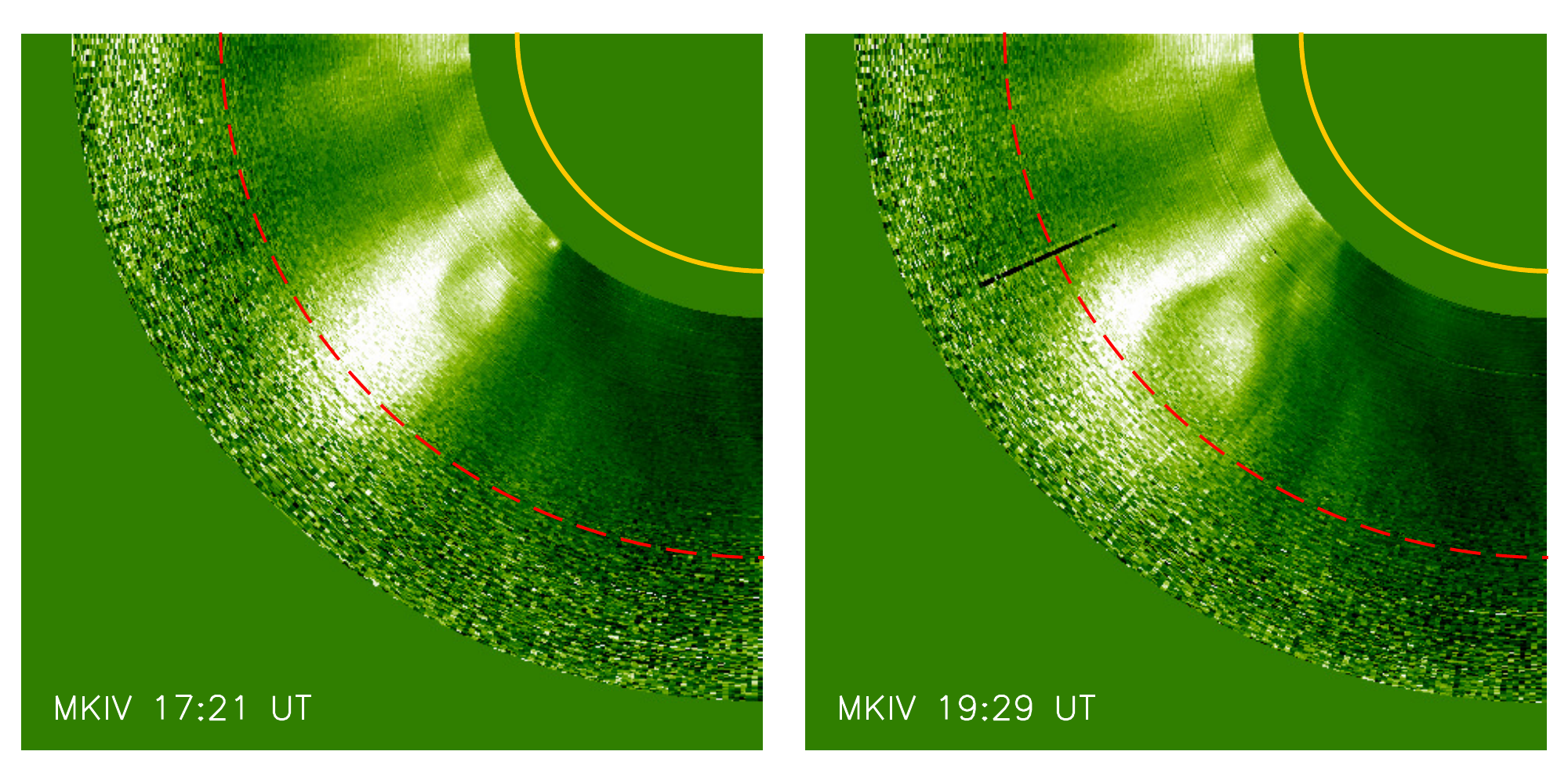}
\end{figure*}

In the event of 2000 November 8 (hereafter Event~1), which is classified as a narrow (angular width of $\sim 59^\circ$), slow CME, the CME leading edge entered in the LASCO-C2 FOV at 19:27~UT with an estimated plane-of-the-sky (POS) speed of $\sim 160$~\kms.
The CME was associated with a prominence eruption and propagated in the South-East quadrant at a latitude of $\sim 30^\circ$S showing the classical, quite symmetrical three-part structure (see Figure~\ref{fig_events}, top row).
The CME launch time, derived by extrapolating the LASCO-C2 measurements to the projected height of 1~\rsun, is approximately at 17:19~UT, in agreement with MLSO/Mark-{\sc iv} image sequences (Figure~\ref{event_1}).
LASCO-C2 acquired a polarimetric sequence for the measurement of the polarized visible light between 20:59 and 21:03~UT.
SOHO/EIT 304~\AA\ and 195~\AA\ images show that the CME probably originated from active region (AR) NOAA 09227 (located at $10^\circ$S $55^\circ$E): prominence material laying above the source AR appears to be ejected between 13:19~UT and 19:19~UT, as showed by the only two EIT 304~\AA\ images available before and after the estimated CME launch time.
The eruption is barely visible in corresponding EIT 195~\AA\ images, where a weak evacuation of plasma followed by a rearrangement of the magnetic configuration above the source AR can be noticed.
Although an M2.9 class flare was detected by the GOES satellite at 16.30~UT -- i.e., very close to the CME start time -- this flare occurred at the NE limb, as shown by TRACE images.

The event of 2000 December 25 (hereafter Event~2) consists in a small-scale eruption of plasma along a coronal streamer triggered by an adjacent CME \citep[see][for similar CME-induced eruptions]{bemporad2010}.
The coronal configuration above the NE limb before the event was characterized by the presence of a faint, quite stable streamer located at a latitude of $\sim 60^\circ$N as shown by the LASCO-C2 white-light images.
The first sign of ejection appeared in the LASCO-C2 FOV at 18:54~UT as a brightening at the visible base of the streamer; the emerging plasma ``blob'' later on traveled along the streamer with an estimated velocity of $\sim 250$~\kms\, showing a complex configuration with evidence of small-scale filamentary structure (see Fig.~\ref{fig_events}, middle row). 
The ejection start time, as extrapolated from the LASCO height-time diagrams, was at 17:34~UT.
The polarimetric sequence for the measurement of the $pB$ was acquired between 20:52 and 21:00~UT.
At 21:04~UT, when the leading edge of the ejected material was above $\sim 4.5$~\rsun, the hemispherical front of a faint CME entered the LASCO-C2 FOV propagating slightly northward to the streamer, at a latitude of $\sim 52^\circ$N, with approximately the same speed (as reported in the LASCO CME catalog).
As the CME progressed, no clear signature of a bright core inside was detected in white light.
EIT 195~\AA\ and 304~\AA\ images show little evidence of an eruption that occurred above the NE limb starting around 17:00~UT, quite consistent with the successive appearance of ejected plasma as seen by LASCO.
The eruption is not associated with flares or EUV brightenings on the disk, hence the source region was most likely located behind the visible disk.

In the event of 2003 May 2-3 (hereafter Event~3), the first appearance of the CME leading edge in the LASCO-C2 FOV was at 23:50~UT on May~2. 
The initial velocity, estimated from a 2$^{\rm nd}$-order polynomial fit, was $\sim 100$~\kms. 
The CME propagated inside a large and bright mid-latitude coronal streamer at the latitude of $\sim 45^\circ$N and exhibited a three-part structure with a faint hemispherical front and clear evidence of an inner core (see Fig.~\ref{fig_events}, bottom row).
At 02:58~UT LASCO-C2 started to acquire the sequence of three polarimetric images for the measurement of the $pB$, until 03:03~UT.
At that time the CME front velocity was around $\sim 220$~\kms.
EIT 304/195~\AA\ images did not show remarkable phenomena occurring below the coronal region where the CME propagated, and the ejection was not associated with a disk event, therefore it most likely originated on the back side of the Sun, probably from ARs 10354/10355, which became visible at the NE limb in EIT~195~\AA\ images on May~4.

\subsection{UVCS data}\label{uvcs_data}

During the three events considered here, the UVCS FOV (a 42~arcmin long slit) performed a radial scan of the coronal region crossed by the CME.
Details about the observations are listed in Table~\ref{table:1}.
Aside from the date and time, we report the radial scan altitude range, the polar angle (PA, measured counter-clockwise from north pole) where the slit was centered, the projected altitude of the slit ($\rho$) at the time of the $pB$ measurement made by LASCO, the exposure time for each spectrum, the projected slit width, and the spatial bin size.
In all cases, the detected spectral range includes the \hi\ \lya\ $\lambda$1215.6~\AA\ line (acquired using the redundant \ovi\ channel with spectral binning of 2 pixels $\simeq 0.1830$~\AA), the \ovi\ $\lambda\lambda$1031.9/1037.6~\AA\ doublet lines (acquired with spectral binning of 1 pixel $\simeq 0.0991$~\AA\ for Event~1 and 2 pixels $\simeq 0.1983$~\AA\ for Events~2 and~3), and other minor lines (not considered in this work).
Standard calibration and data reduction was made with the latest version available of the UVCS Data Analysis Software (DAS 51).

\begin{table*}[tb]
\footnotesize
\caption{Summary of UVCS observations.\label{table:1}}
\begin{tabular}{ccccccccc}
\tableline\tableline

Event & Date & Time\tablenotemark{a} & Altitude range & PA    & $\rho$\tablenotemark{b}  & $t_{\rm exp}$\tablenotemark{c} & Projected slit width & Spatial bin size \\
      &      &                       & (\rsun)        & (deg) & (\rsun)                  & (s)                            & (arcsec)   & (arcsec) \\
\tableline
1 & 2000 Nov 08 & 18:02~UT & $1.5-4.0$                  & 113 & 2.5 & 300  & 28 & 42 \\
2 & 2000 Dec 25 & 20:58~UT & $1.6-3.0$\tablenotemark{*} & 45  & 3.0 & 120  & 42 & 21 \\
3 & 2003 May 02 & 02:48~UT & $1.6-3.5$\tablenotemark{*} & 45  & 3.5 & 120  & 42 & 21 \\
\tableline

\end{tabular}
\tablenotetext{a}{Initial time of UVCS observations.}
\tablenotetext{b}{Heliocentric distance of the UVCS slit at the time of LASCO-C2 $pB$ measurement.}
\tablenotetext{c}{Exposure time.}
\tablenotetext{*}{For this event, the radial scan was performed from higher to lower heights.}
\end{table*}

In what follows, our analysis focuses on the \hi~\lya\ intensities and profiles.
For each event, we averaged the two exposures acquired around the time of the LASCO $pB$ measurement, in order to increase the signal-to-noise ratio; for Event~2 and~3 we also averaged the data over two spatial bins ($=42$~arcsec).
We then fitted a single Gaussian to the observed \lya\ spectrum at each position along the slit, getting line intensities and widths.
Usually spectra in the optically thin corona must be corrected for the ambient-corona contribution before spectral integration, in order to isolate the CME emission with respect to the emission of the surrounding corona aligned with the CME along the same line of sight. However, in all the three cases, observations by UVCS started when the CME was already crossing the instrument FOV, making impossible to get a reliable pre-event \lya\ spectrum to be subtracted from the CME exposures to remove the quiet-corona contribution.
To account for that, after integration of the \lya\ spectra, we modeled the quiet-corona intensity at each position along the slit using a power-law extrapolation (with a $r^{-\alpha}$ dependence on the heliocentric distance), constraining the fit to the intensity measured in the slit regions not crossed by the CME. The intensity from the fit extrapolated at the spatial region crossed by the CME was thus subtracted from the measured one.

The line widths $\Delta\lambda_{1/e}$ obtained from the Gaussian fit to the \lya\ line profiles were corrected for the instrumental profile broadening and converted into effective temperatures with the usual relationship
\begin{equation}
\Delta\lambda_{1/e}=\frac{\lambda_0}{c}\sqrt{\frac{2k_BT_{\rm eff}}{m_H}},
\end{equation}
where $\lambda_0$ is the spectral line central wavelength at rest, $c$ is the light speed, $k_B$ the Boltzmann constant, $m_H$ the hydrogen mass, and $T_{\rm eff}$ is the so-called effective temperature.
Note that the observed profiles are likely broadened by non-thermal motions as well as by the line-of-sight component of the plasma bulk velocity; moreover, since we were not able to subtract a pre-event spectrum, they include a component relevant to the quiet corona as well.
For these reasons, the derived effective temperatures are only an upper-limit estimate of the real hydrogen kinetic temperature.

The additional analysis of the \ovi\ $\lambda\lambda 1032/1037$~\AA\ doublet lines detected by UVCS could in principle provide further information on the CME plasma outflow velocity. In fact, the \ovi\ intensity ratio, $R=I_{1032}/I_{1037}$, is sensitive to the radial component of the plasma outward speed through the Doppler-dimming effect \citep[see, e.g.,][]{noc87}; the outflow velocity can be thus inferred from the comparison of the measured ratios with those predicted by models.
In general, values of $R \gtrsim 2$ imply outflow velocities around or lower than 100~\kms, while at higher speeds, above $\sim 150$~\kms, pumping of the \ovi\ $\lambda1037$~\AA\ by the {\sc C ii} $\lambda\lambda 1036/1037$~\AA\ lines causes $R$ to be lower than 2 \citep[see, e.g.,][]{ray04,cia05}.
Moreover, the dependence of $R$ on the outflow velocity is function of the electron density and temperature \citep[see][]{dob03}.

Nevertheless, two reasons led us to neglect results based on the analysis of the \ovi\ line ratios.
First, the uncertainties on the \ovi\ line intensities are larger than those affecting the \lya\ line intensity, due to lower count statistics; moreover, since we were not able to evaluate correctly the coronal background intensity, we have larger uncertainties in the measurement of $R$ (of the order of $\sim 30$\% for Event~1 and $\sim 50$\% for Event~2 and~3).
The consequence is that in all the three cases, the intensity ratio does not differ significantly from 2 in most part of the UVCS FOV \citep[according to][we considered as statistically significant only departures of the ratio from the value of 2 larger than $2\sigma$]{ray04}.
Second, as we will discuss in the next sections, the analysis of the white-light emission measured by LASCO-C2 provided electron densities that are of the order or larger than $10^6$~cm$^{-3}$ in all cases.
As clearly demonstrated by \citet{dob03} (for the 1999 April 15 CME at 1.98~\rsun), the \ovi\ intensity-ratio at typical coronal temperatures and for electron densities higher than $10^6$~cm$^{-3}$ is $R\simeq 2$ in a wide range of velocities, so that it is impossible to identify a unique value for the CME outflow speed owing those conditions.
Therefore, we could not place reliable constraints on this parameter using these results and we preferred to estimate the outflow plasma velocity using the analysis of the CME dynamics in LASCO-C2 VL images (see below).

\section{Diagnostic methods}

\subsection{Density diagnostics from polarized visible light}

The coronal visible light (K-corona) is produced by the Thomson scattering of the photospheric incident radiation off coronal free electrons; VL images can provide the electron column density $N_e = \int_{\rm los} n_e\,dz$ using the method described, for instance, by \citet{vou00}: the number of scattering electrons along the LOS (i.e., the column density $N_e$, in units of cm$^{-2}$) is given by the ratio of the observed total brightness $B$ over the brightness $B_{\rm exp}(z)$ expected for a single electron located at distance $z$ from the plane of the sky.
Usually this computation is done pixel-by-pixel in the CME exposure under the assumption that the scattering electrons are located on the plane of the sky ($z=0$).
In this way, a lower limit estimate of the column density $N_e$ is obtained.
However, in their quantitative comparison between the real 3D structure of a flux-rope CME obtained from MHD simulations and that derived via polarization-ratio technique, \citet{pag15} recently demonstrated that if $B_{\rm exp}$ is calculated by assuming $z=\avg{z_{\rm cme}}$ with $\avg{z_{\rm cme}}$ center of mass of the LOS plasma distribution, the resulting value of $N_e$ is much closer to the real one, with uncertainties generally lower.

An estimate of the location of the emitting plasma along the LOS, $\avg{z_{\rm cme}}$, can be obtained with the so-called polarization-ratio technique, first proposed by \citet{mor04} to infer from a single-viewpoint $pB$ image of a CME the three-dimensional structure of the ejected plasma.
The degree of polarization introduced by the Thomson scattering is function of the scattering angle between the incident photon direction and the direction towards the observer \citep[see][]{bil66}, and, in turn, of the distance $z$ of the scattering electron from the POS. As shown by \citet{mor04}, the dependence of the ratio $p=pB/B=p(z)$ for a single electron on the distance from the plane of the sky can be exploited to determine, pixel by pixel in the CME image, the mean position $\avg{z_{\rm cme}}$ of the CME plasma. The more the region where the electron density is locally increased by the CME is spatially limited along the LOS, the more this determination is accurate \citep[see][]{bem15}.

On the other hand, owing to the forward/backward symmetry of the Thomson scattering process, it is impossible to establish with this method if the emitting plasma is actually located in front of or behind the plane of the sky, because $\avg{z_{\rm cme}}$ can be derived aside from its sign. Hence, additional considerations are needed to discriminate between these two possible cases. A key point in applying the polarization-ratio technique is the coronal background removal, which is necessary to exclude from the observed emission contributions coming from both the surrounding ambient corona and the dust-scattered (F-corona) component that affects the unpolarized part of the total brightness.
These contributions, that sum together with the genuine CME emission because of the optical thinness of the solar corona, would result in an erroneous determination of the CME 3D structure.

The background removal is usually performed by subtracting a convenient pre-CME image from the frame that is analyzed; for the polarized visible light, this operation can be achieved in several ways: for instance by subtracting the pre-event Stokes parameters from the Stokes images of the CME, or by subtractig pre-event $B$ and $pB$ images directly from the corresponding $B$ and $pB$ images of the CME \citep[see, e.g.,][]{dere05}, or even by subtracting pre-event frames separately from all the polarized images before they are combined to obtain the $pB$ \citep[see][]{mor04}.
On the other hand, \citet{mie09} suggested a different approach that consists in subtracting from the exposure containing the CME a minimum-intensity image instead of a single pre-event frame. The minimum-intensity image is created by taking, pixel by pixel, the minimum brightness value over a sequence of exposures acquired before and during the event.

In our analysis, we followed the methods described in \citet{mor04} and \citet{mie09} within certain limitations.
As stated above, LASCO-C2 usually acquires one or two sequences of three polarimetric images per day.
With this quite poor temporal cadence, the minimum-intensity background image cannot be created with more than one pre-CME exposure, because the configuration of the ambient corona two days before each event can be different from that existing during the CME.
For this reason, after standard calibration of the VL images \citep[which includes correction for vignetting and offset bias, and the radiometric calibration; see][for further details about the LASCO-C2 calibration]{llebaria08}, we computed a minimum-intensity image (for each orientation of the C2 polarizers: $0^\circ$, $+60^\circ$, and $-60^\circ$) using the exposures acquired soon before, during, and just after each event.
In this way, the removal of steady structures not affected by the transit of the CME, such as streamers, turns out to be better than considering only the pre-event and the event images.
In the specific case of Event~3, we also took into account the different calibration of the CME images that were acquired with the ``blue'' filter of the LASCO-C2 instrument, unlike all the other images considered in this work, which were taken with the ``orange'' one.

Another issue to be accounted for is the effect of the CME motion between the polarimetric exposures, which can be source of additional uncertainties in the determination of the polarized brightness because of the partial misalignment between the same CME structures in different images.
This is a minor concern in our case, however, because the events analyzed here were slow or very slow CMEs: we estimated that in the worst case (i.e. Event~2, where the maximum projected velocity was $\approx 250$~\kms) the CME front moved less than 2 pixels between consecutive exposures.
In order to compensate for this displacement, and following the same procedure adopted by \citet{mor04}, we smoothed each polarized exposure using a 3-pixel-by-3-pixel median filter.
In this way, information at spatial scales equal or lower than 3 pixels ($=71.4$~arcsec) is lost, but this also significantly reduces the errors associated with the CME motion.

\begin{figure*}
\caption{\label{polarization}{\em Left column:} polar maps (as functions of heliographic latitude and heliocentric distance) of the polarized visible light measured by LASCO-C2 for Event~1 (top row), Event~2 (middle row), and Event~3 (bottom row), after background subtraction and processed with the NRGF. {\em Middle column:} topographical maps of the average location along the LOS of the CME plasma, $\avg{z_{\rm cme}}$. The white areas correspond to coronal regions excluded from the calculations, as described in the text. {\em Right column:} same, for the electron column density $N_e$ derived from the $pB$. In all panels, the UVCS slit is superimposed as a red line. The solar disk (yellow area) is reported for reference purposes.}
\centering
\includegraphics[width=\textwidth]{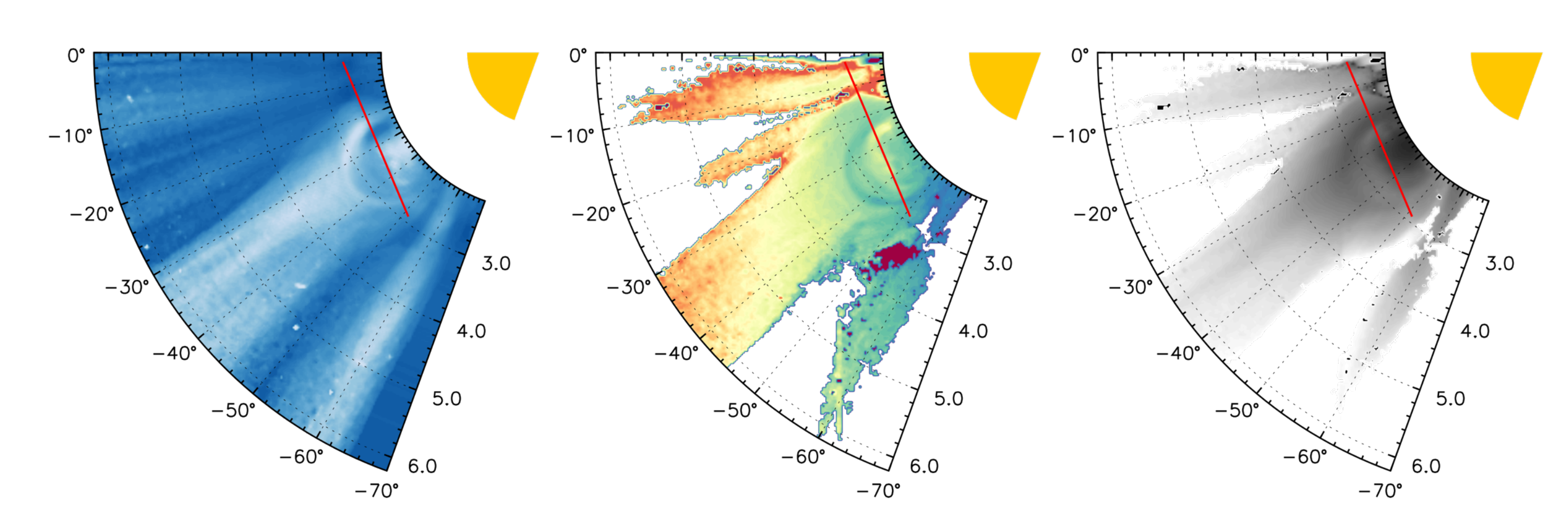}
\includegraphics[width=\textwidth]{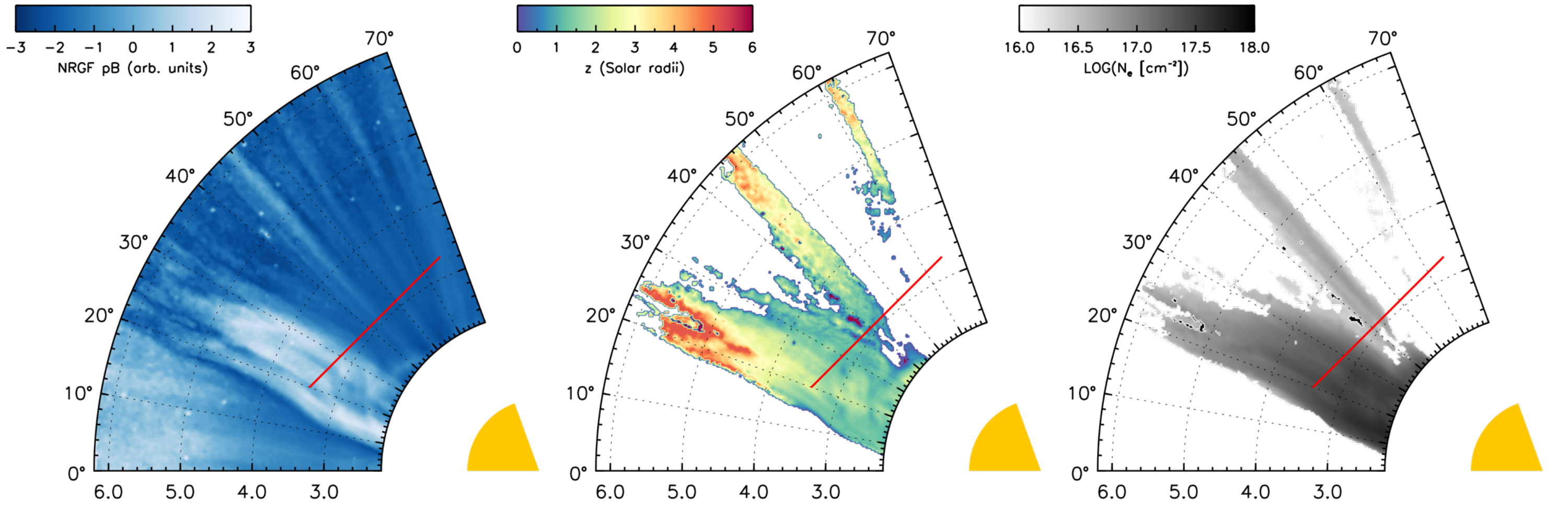}
\includegraphics[width=\textwidth]{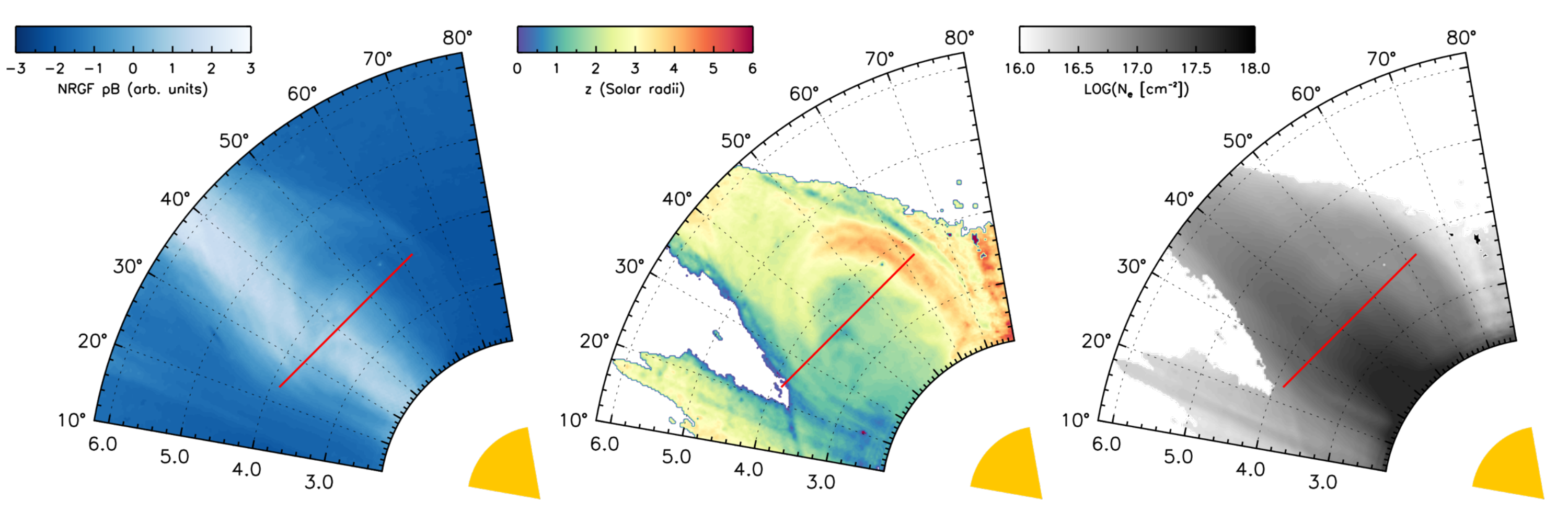}
\end{figure*}

The polarimetric images were then used to compute the Stokes parameters ($I$, $Q$, and $U$) that were corrected for the small instrumental elliptical polarization produced by the two folding mirrors in the beam path, which was removed with use of the M\"uller matrix formalism following the method described in \citet{mor06}.
Finally, the $B$ and $pB$ images of the CME were obtained (see Figure~\ref{polarization}, left column) with the standard relationships $B\equiv I$ and $pB\equiv\sqrt{Q^2+U^2}$ \citep[see][]{lan04}. In this way the radiometric calibration of the polarized exposures is not strictly necessary when applying the polarization-ratio technique, because the ratio $pB/B$ is used to determine the position of the emitting plasma along the LOS; conversely, it is essential for the derivation of the electron density (see below). As explained by \citet{frazin12}, the uncertainty in the standard radiometric calibration of C2 can be considered reasonably within $\pm15$\% of the measured $pB$ and $B$ values, so we used this value for the estimate of the errors associated with all the other derived quantities.

For each pixel in the background-subtracted CME $pB$ image, the mean distance from the POS of the plasma responsible for the emission in that pixel was estimated by comparing the observed ratio $p_{\rm obs}$ with the theoretical ratio $p_{\rm exp}$ expected for a single electron, according to the Thomson-scattering theory.
We used the SolarSoft routine {\tt eltheory.pro} to compute the values of $p_{\rm exp}$ for a range of distances $z$ from the POS and found the value $\avg{z_{\rm cme}}$ for which $p_{\rm obs}=p_{\rm exp}$.
Note that, as shown by \citet{bem15}, the resulting $\avg{z_{\rm cme}}$ actually represents the location along the LOS of the center of mass of a ``folded'' density distribution, given by reflecting and summing in front of the POS the fraction of the real density distribution located behind the POS.
This leads to over/underestimates of the real distance from the POS of the emitting plasma and the derived $\avg{z_{\rm cme}}$ can be considered more accurate only when the emitting plasma is at an angle of $\sim 20^\circ$ from the plane of the sky.

The topographical $z$ maps derived for the three events are shown in Figure~\ref{polarization} (center column) together with the background-subtracted $pB$ maps (filtered using the NRGF filter to highlight the latitudinal structures; left column).
Note that the calculation of $\avg{z_{\rm cme}}$ was restricted only to pixels with values of $pB$ and $B$ above 10\% of the mean, in order to prevent artifacts in the 3D reconstruction.

The map relevant to Event~1 shows that the CME front and core plasma is concentrated at an average distance of~$\sim 1.6$~\rsun\ from the POS (i.e., at an average angle of $\sim 30^\circ$), while in the regions surrounding the front and in the cavity (the low-brightness region inside the CME) the emission appears to be associated to plasma located at greater distances from the POS.
This result can be interpreted in the light of what is found by \citet{bem15}: the density distribution of the plasma in the void of the CME and in the surrounding quiet regions is most probably extended along the LOS more than that in the front and in the core, which are more spatially confined; therefore, in the former regions the tail of the density distribution behind the POS is reflected back in front of the POS when the polarization-ratio technique is applied (due to the sign ambiguity of the method, as explained above) and this leads in turn to an overestimate of $\avg{z_{\rm cme}}$.

The $z$-map of Event~2 shows that around 21:00~UT the leading edge of the ejected plasma, located at a heliocentric distance of $\sim 4.5$~\rsun, had correspondingly reached the maximum average distance from the POS of $\sim 3.5$~\rsun, consistent with a blob of plasma travelling in a direction forming an angle of $\sim 40^\circ$ from the POS.
At lower altitudes, other small-scale structures located approximately along the same direction with respect to the plane of the sky can be recognized in the map.
Note that, as in the previous case, the brightest features identifiable in the $pB$ map are located systematically closer to the POS in the $z$-map.

Concerning Event~3, the CME plasma appears to be located at an average distance of about 1.5~\rsun\ from the POS in the southern leg of the front, while at higher distances, $> 3$~\rsun, in the northern one.
Note that the southern part of the CME front was mostly superimposed to the pre-event coronal streamer, as evidenced by LASCO total-brightness images, so the difference in the position along the LOS between the two legs could be due to the contribution of the spurious white-light emission coming from the streamer. 

\begin{figure}
\caption{\label{densities} Electron densities (grey shaded areas) and UVCS \hi\ \lya\ intensities (heavy line) as functions of the heliographic latitude along the UVCS field of view. Electron densities were derived using two values of the thickness of the CME along the LOS: $L=0.25$~\rsun (upper edge of the shaded area) and $L=0.5$~\rsun (lower edge), as described in the text.}
\centering
\includegraphics[width=0.5\columnwidth]{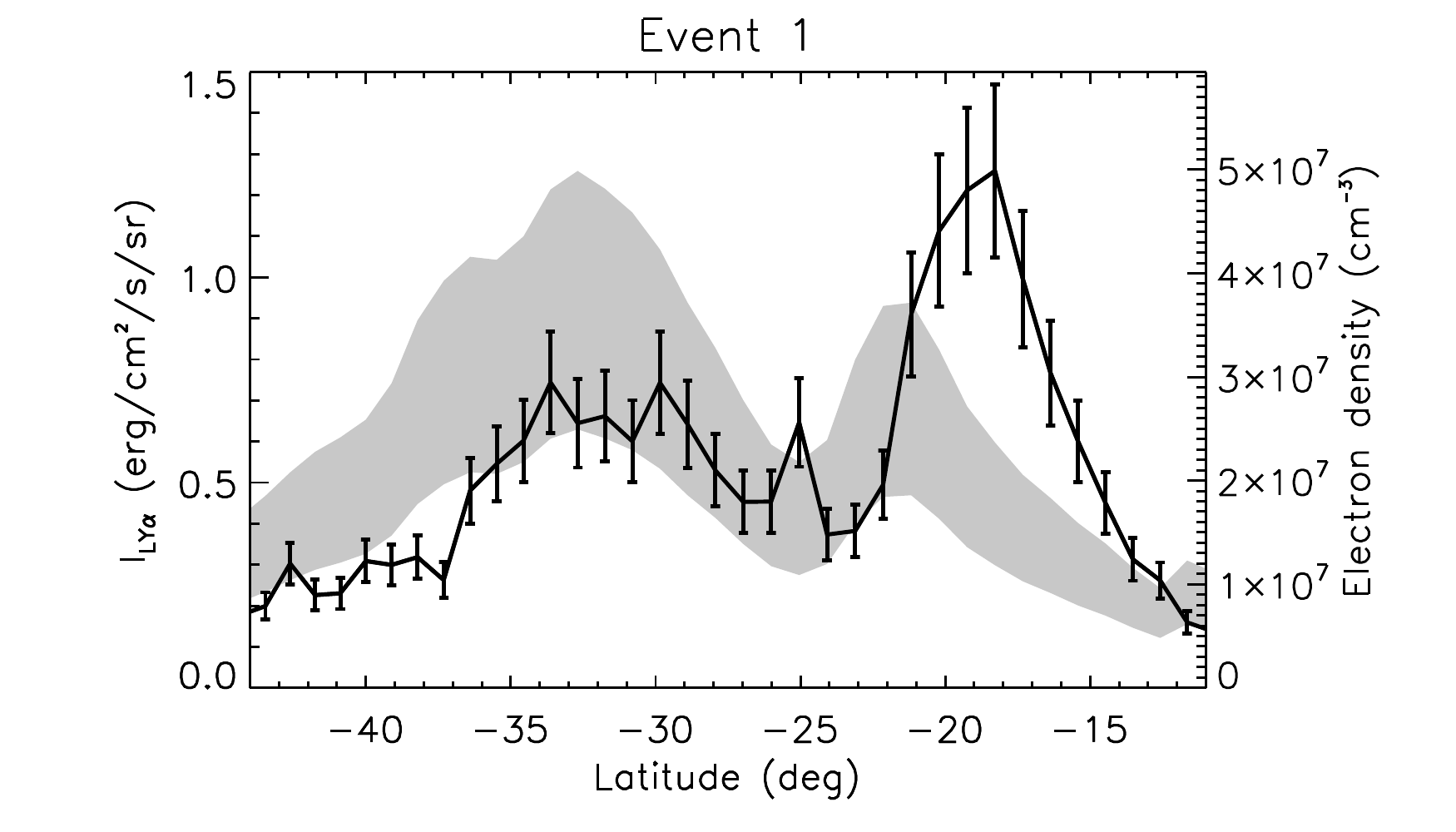}
\includegraphics[width=0.5\columnwidth]{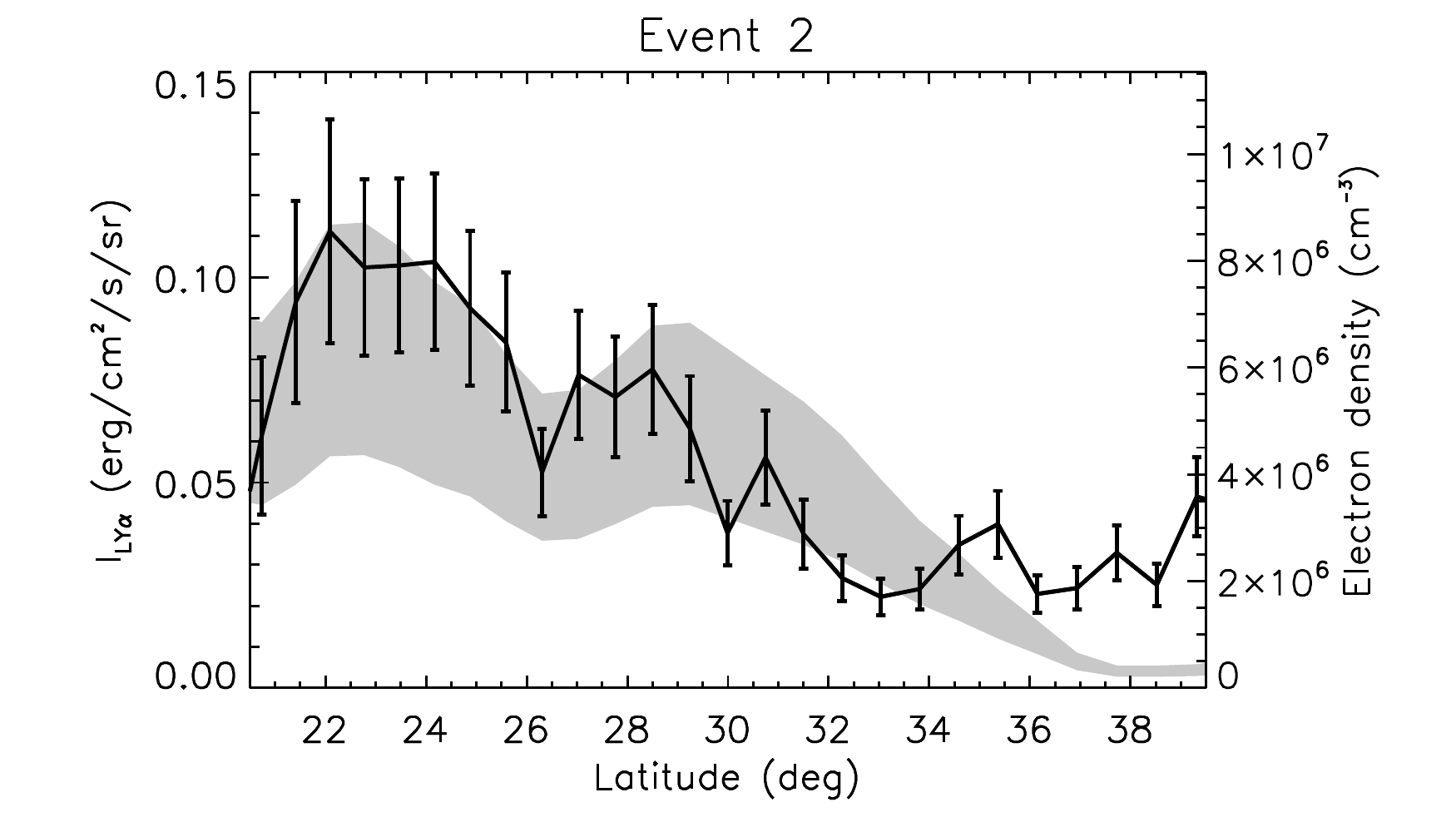}
\includegraphics[width=0.5\columnwidth]{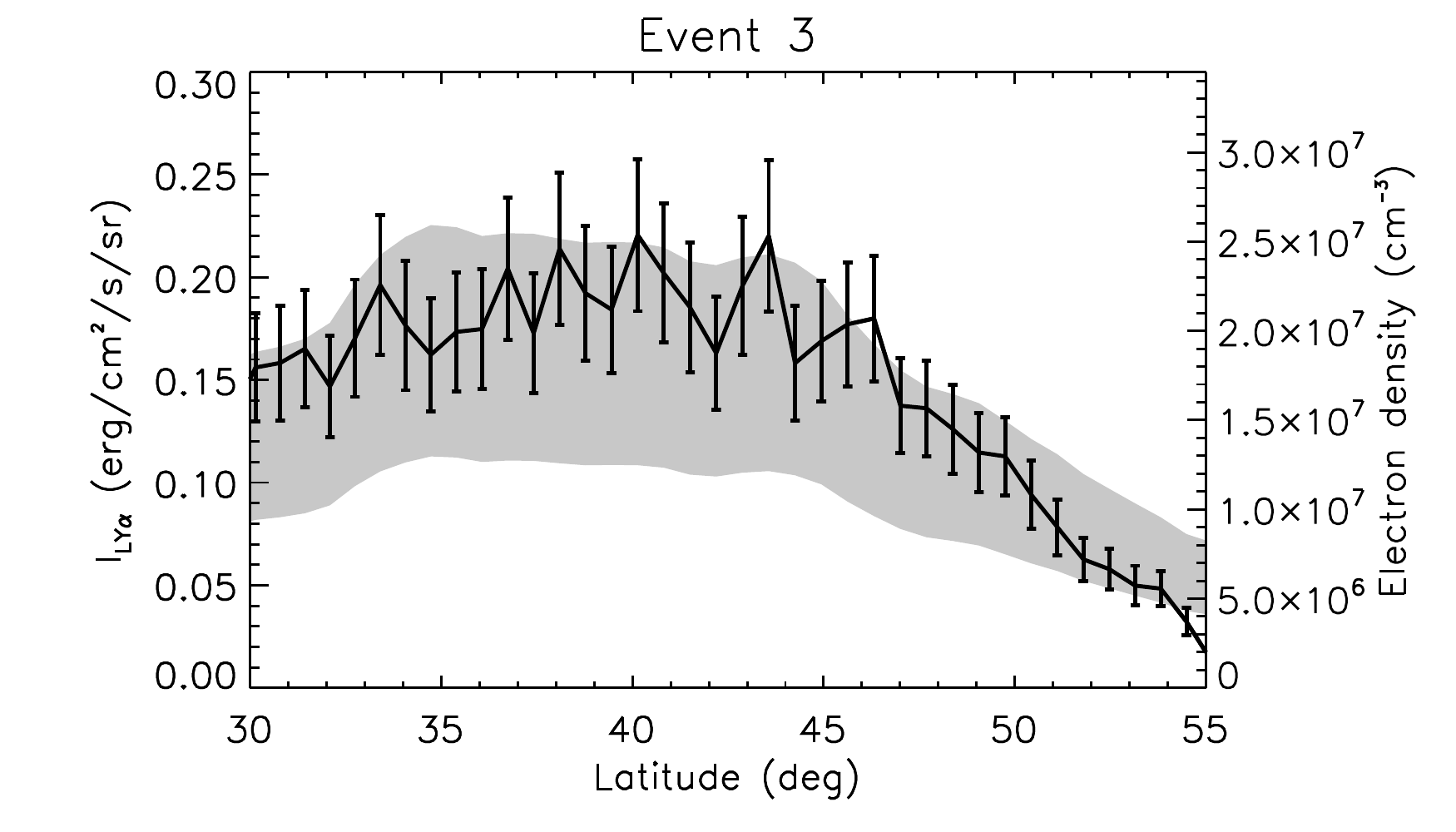}
\end{figure}

Finally, the resulting $z$-maps of LOS plasma distribution were used to derive the corrected column density maps \citep[see][]{pag15}, shown in Figure~\ref{polarization} (right column). 
The column density can be then converted to volume density (units of cm$^{-3}$) if the thickness $L$ of the CME plasma along the LOS is known; in fact, it is
\begin{equation}
N_e\equiv\int_{\rm los}{n_e\,dz}\approx\avg{n_e}\cdot L
\end{equation}
where $\avg{n_e}$ is the average electron density of the CME plasma \citep[see also][for a discussion on the derivation of electron density maps from the inversion of VL images.]{quemerais02}.
Unfortunately, the parameter $L$ cannot be directly estimated from VL images, nor derived with the polarization-ratio technique, but it must be reasonably assumed.
For each of our events, we derive the average electron density assuming for $L$ the two values of 0.25 and 0.5~\rsun.
We report in Figure~\ref{densities} the density profiles interpolated along the UVCS slit, together with the corresponding \lya\ intensity profiles derived with UVCS.

The electron densities obtained in the three cases are quite consistent with those usually measured in the core of CMEs with other techniques \citep[$\sim 10^6-10^7$~cm$^{-3}$; see, e.g.,][]{cia00,cia01,dob03,bem07}.
Event~2, which turns out to be the faintest in white light, accordingly exhibits the lowest densities (up to $\sim 8\times 10^6$~cm$^{-3}$).
Event~3 shows a noticeable latitudinal correlation between the distribution of the electron density and that of the \lya\ intensity.
Conversely, for Event~1 and~2 the locations where the density has local maxima appear to be less correlated with the \lya\ intensity peaks, in particular those located around $29^\circ$N for Event~2 and around $20^\circ$S for Event~1.

We used the electron densities derived from the analysis of LASCO VL images to evaluate the \lya\ line intensities expected along the UVCS FOV in the three cases; comparison with the intensities actually measured by UVCS have been used to constrain possible ranges of electron temperatures of the CME plasma under certain assumptions, as we will discuss in the following.

\subsection{Combined analysis of visible light and \lya\ UV radiation}

The mechanism of formation of many strong coronal lines in the UV (such as the \hi\ \lya\ or the \ovi\ doublet lines) is a combination of radiative and collisional excitations, followed by spontaneous emission. The total intensity of such lines is then the sum of a radiative and a collisional component, $I=I_{\rm rad}+I_{\rm col}$. The radiative component is produced by resonant scattering of the chromospheric radiation by coronal ions and can be affected by the outflows of the scattering atoms through the Doppler-dimming effect.
We refer the reader to \citet{koh82} and \citet{noc87} for a complete treatment of the resonant scattering and Doppler dimming processes in coronal plasmas.

An approximate expression for the radiative component, holding for coronal structures that are spatially limited along the LOS \citep[e.g., coronal streamers or CMEs; see also][]{koh06}, is
\begin{equation}\label{irad}
I_{\rm rad}\simeq\frac{h}{16\pi} B_{12}\cdot b\cdot \lambda_0\cdot \Omega\cdot F_D(v_{\rm out})\int_{\rm los}{n_i\,dz},
\end{equation}
where $h$ is the Planck constant, $B_{12}$ the Einstein coefficient for absorption, $b$ the branching ratio for radiative de-excitation, $\lambda_0$ the wavelength of the line transition, $\Omega$ the solid angle subtended by the solar disk at the scattering location, $n_i$ the ion number density, and
\begin{equation}\label{dimming}
F_D(v_{\rm out})=\int_{0}^{\infty}{I_\odot(\lambda-\delta\lambda)\Phi(\lambda-\lambda_0)\,d\lambda}
\end{equation}
is the so-called Doppler-dimming factor, which is function of the intensity line profile of the chromospheric radiation, $I_\odot(\lambda-\delta\lambda)$ -- shifted by the quantity $\delta\lambda=(v_{\rm out}/c)\cdot \lambda_0$, i.e., the Doppler shift introduced by the radial component of the plasma outflow velocity $v_{\rm out}$ -- and the normalized coronal absorption profile along the direction of the incident radiation, $\Phi(\lambda-\lambda_0)$.

The collisional component is due to de-excitation of a coronal ion previously excited by collision with a free electron. 
Approximately it is given by
\begin{equation}\label{icol}
I_{\rm col}\simeq\frac{1}{4\pi}b\cdot q_{\rm col}(T_e)\int_{\rm los}{n_e\cdot n_i\,dz},
\end{equation}
where the term
\begin{equation}
q_{\rm col}(T_e)=2.73\times 10^{-15}\cdot \bar{g}\cdot f_{12}\cdot\frac{1}{{E_{12}}}\cdot\frac{1}{\sqrt{T_e}}e^{-\frac{E_{12}}{k_BT_e}}
\end{equation}
is the collisional excitation rate, which depends on the average Gaunt factor, $\bar{g}$, the oscillator strength of the transition, $f_{12}$, the transition energy, $E_{12}=hc/\lambda_0$, and the electron temperature, $T_e$.

In both Equations~(\ref{irad}) and~(\ref{icol}), the numerical density of the emitting ions can be approximated using the relationship \citep[see, e.g.,][]{wit82}
\begin{equation}
n_i\approx 0.83\cdot A\cdot R(T_e)\cdot n_e,
\end{equation}
where the factor 0.83 is the ratio of proton to electron density in the case of fully ionized plasma with 90\% of H and 10\% of He (as in typical coronal plasma conditions), $A$ is the element abundance relative to hydrogen ($A=1$ for H$^{\rm 0}$ ions), and $R(T_e)$ is the element ionization fraction, which is function of the electron temperature.
Therefore, if $L$ is the thickness along the LOS of the emitting plasma, the above equations simplify to:
\begin{align}
I_{\rm rad} & \propto R(T_e)\cdot F_D(v_{\rm out})\cdot\avg{n_e}\cdot L\label{comp1}\ \rm{and}\\
I_{\rm col} & \propto R(T_e)\cdot q_{\rm col}(T_e)\cdot\avg{n_e}^2\cdot L.\label{comp2}
\end{align}
Equations~(\ref{comp1}) and~(\ref{comp2}) can be applied to the case of the \hi\ \lya\ emission and used to constrain the electron temperature by the excitation rates needed to account for the observed line intensity.

\begin{figure}
\caption{\label{model} Modeled radiative (solid line) and collisional (dashed line) components of the \hi~\lya\ intensity as functions of the electron temperature $T_e$, for two representative values of the the plasma outflow velocity (top and bottom panels, respectively), three values of the electron density (marked with different line thickness), and two values of the assumed thickness of the emitting plasma volume along the LOS (the upper line in each couple corresponds to the greater value of $L$, see the text).}
\centering
\includegraphics[width=0.5\columnwidth]{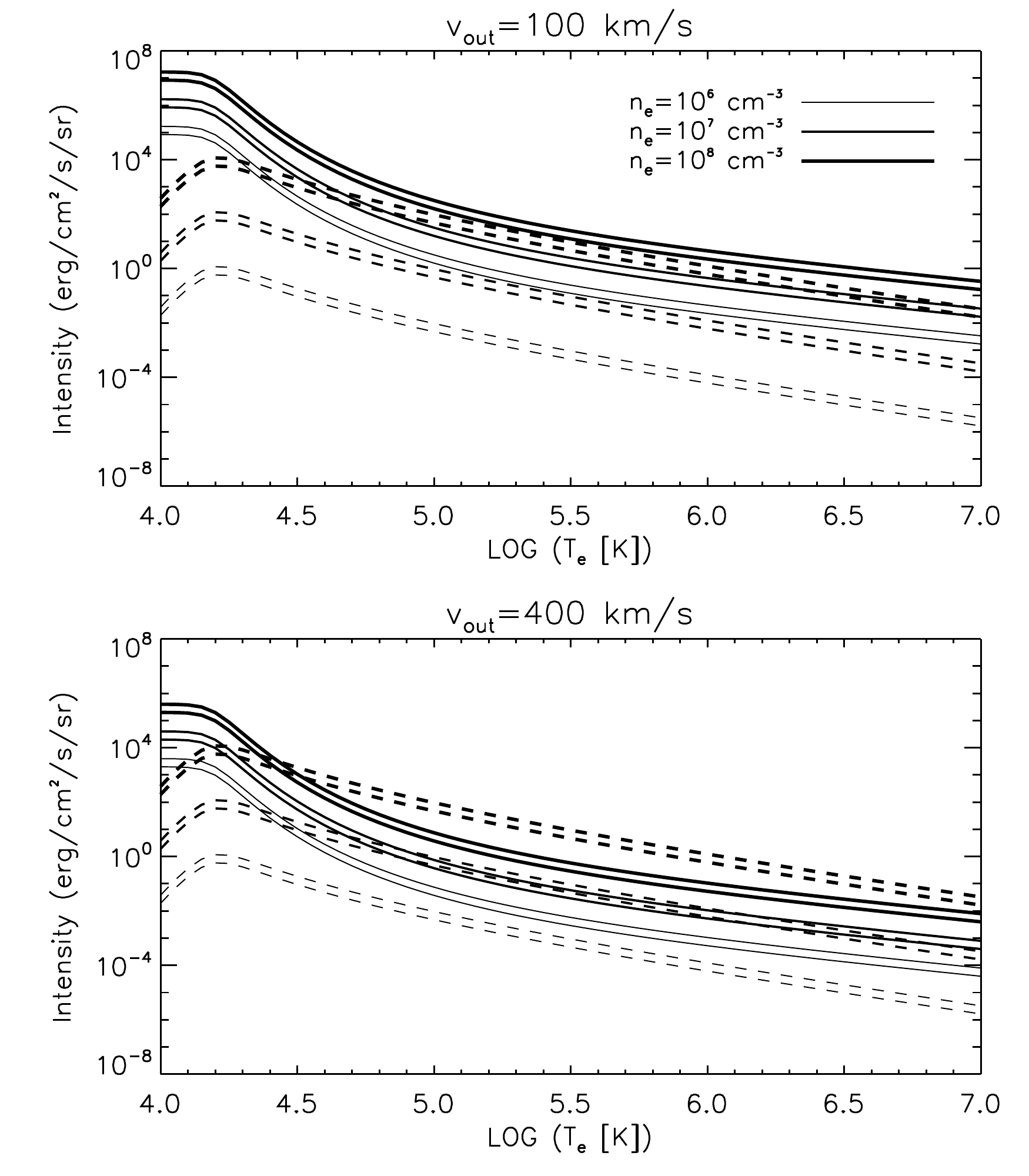}
\end{figure}

As an example, we report in Figure~\ref{model} the prediction of the radiative and collisional components of the \hi~\lya\ intensity as functions of the electron temperature $T_e$, for two representative values of the the plasma outflow velocity (100 and 400~\kms), three values of the electron density ($10^6$, $10^7$, and $10^8$~cm$^{-3}$), and two values of the thickness along the LOS ($L=0.25$~\rsun\ and $L=0.5$~\rsun).
These estimates were made using standard values for all the atomic coefficients and the other quantities entering Equations~(\ref{comp1}) and~(\ref{comp2}), and by adopting the ionization equilibrium provided by the CHIANTI atomic database (version 7) for hydrogen. The idea of our analysis is to derive the expected total \hi\ \lya\ intensity $I_{\rm exp}(T_e)$, as a function of $T_e$, given by $I_{\rm exp}(T_e)=I_{\rm rad}(T_e)+I_{\rm col}(T_e)$ and then to determine $T_e$ from a comparison between $I_{\rm exp}(T_e)$ and the real observed intensity.

Nevertheless, the relative importance of radiative and collisional contributions is related also to the plasma outflow speed. For the \lya\ transition, the radiative component is usually dominant in coronal plasmas; however, at low velocities, due to the Doppler-dimming effect, it becomes comparable to the collisional one at coronal temperatures ($10^5-10^6$~K) when the electron density is higher than $10^8$~cm$^{-3}$. At higher velocities, the radiative component is significantly dimmed, and it can be even lower than the collisional one when the electron density exceeds $10^7$~cm$^{-3}$.
It is worth noting that the variation of the LOS thickness $L$ of the emitting plasma has a much lower effect on the \lya\ intensity than the variation of the outflow velocity and/or the electron density.

Therefore, the determination of CME plasma temperatures requires the constraint of the outflow plasma velocity as well, in order to estimate the Doppler dimming factor $F_D$. A first estimate is simply provided by the POS component of the CME speed, $v_{\rm pos}$, measured with LASCO.
We used the two consecutive total-brightness exposures acquired before and after the time of the $pB$ measurement, to determine the radial component of the CME front projected on the POS, by locating homologous points along the front at the two different times and computing $v_{\rm pos}=\Delta h/\Delta t$. This allowed us to derive the profile of the radial velocity in different points located along the UCVS slit FOV (i.e., at different latitudes).
Nevertheless, the $v_{\rm pos}$ is an underestimate of $v_{\rm out}$ (which is the real radial component of the plasma outflow velocity) because of the projection effect.
It is possible to correct for this effect and obtain a better estimate of real $v_{\rm out}$ by using information on the angle $\varphi$ between the POS and the reconstructed position of a CME plasma element derived with the polarization-ratio technique, owing the relationship 
\begin{equation}
\cos\varphi=\frac{\rho}{\sqrt{\rho^2+\avg{z_{\rm cme}}^2}}.
\end{equation}
Then for any position along the UVCS slit it is $v_{\rm out}=v_{\rm pos}/\cos\varphi$; this also shows how the polarization-ratio technique can be used to derive from single view-points (as those that will be provided with Metis) the real unprojected speed and CME propagation direction, two very important parameters to help constraining the CME propagation time and for Space Weather forecasting purposes.

Once the unprojected velocity has been determined, it can be used to estimate the corresponding Doppler-dimming factor $F_D(v_{\rm out})$; the assumptions we made for the calculation of this parameter (entering Equation~\ref{comp1}) deserve some considerations. The choice of the incident radiation profile from the lower atmosphere, $I_\odot(\lambda)$, is crucial; we adopted here the \lya\ disk profile reported by \citet{lem02} and measured with SOHO/SUMER on November 12, 2000, i.e., very close to the dates of Event~1 and~2; note that this profile was reconstructed with a maximum under-estimation of 1\%, as stated by the authors.
We used the same line profile also for Event~3 because, although the integrated \lya\ flux may vary by 80\% between the solar maximum and minimum \citep[e.g.,][]{tob97}, the spectral line profile does not appear to change significantly over the solar cycle \citep[see][]{lem02}.
The line profile was scaled so that the total integrated line intensity matched the \lya\ irradiance measured by the UARS/SOLTICE \citep[][]{woo00} instrument at the times of Events~1 and~2 ($5.34\times 10^{11}$ and $5.94\times 10^{11}$~photons~cm$^{-2}$~s$^{-1}$, respectively) and by the TIMED/SEE instrument at the time of Event~3 ($4.70\times 10^{11}$~photons~cm$^{-2}$~s$^{-1}$).
Note that these values are more than a factor of $\sim 2$ higher than the average quiet-Sun \lya\ flux reported by \citet{ver78}, which is often used as a reference; this difference is consistent with the fact that the value reported in \citet{ver78} was measured during the solar minimum, while our events are close to the maximum.

The absorption profile $\Phi(\lambda-\lambda_0)$ was assumed to be Gaussian with a $1/e$ line width equal to the average \lya\ line width measured by UVCS in the three cases.
Since the measured line width is actually an upper limit, because of the possible sources of line broadening that have been neglected, the calculated Doppler-dimming factor is underestimated, and this leads in turn to an underestimate of the radiative component of the \lya\ line.

Once the Doppler-dimming factors are computed, given the average electron densities derived from LASCO in the specific cases of our events, we evaluated for each position along the UVCS FOV, the radiative and collisional component of the \lya\ line for a range of electron temperatures, and obtained the expected total intensity as function of the temperature, $I_{\rm exp}(T_e)$. We then found through inspection the temperature at which the expected intensity matches the observed one. In this way, we determined at the same time not only the electron temperatures $T_e$, but also the relative contributions of the radiative and collisional components needed to reproduce the observed total intensity.

\section{Resulting CME electron temperatures}

\begin{figure}
\caption{\label{temps} \hi\ \lya\ effective temperature derived from UVCS spectra (red shaded area) and electron temperature derived from the analysis described in the text (blue shaded area), as functions of the heliographic latitude along the UVCS field of view. The vertical width of the areas is equal to the uncertainties affecting the results. The dashed and dotted lines represent the electron temperature derived with the radiative and collisional approximations, respectively (see the text). The location of the CME core is indicated in the plots relevant to Events~1 and~3.}
\centering
\includegraphics[width=0.5\columnwidth]{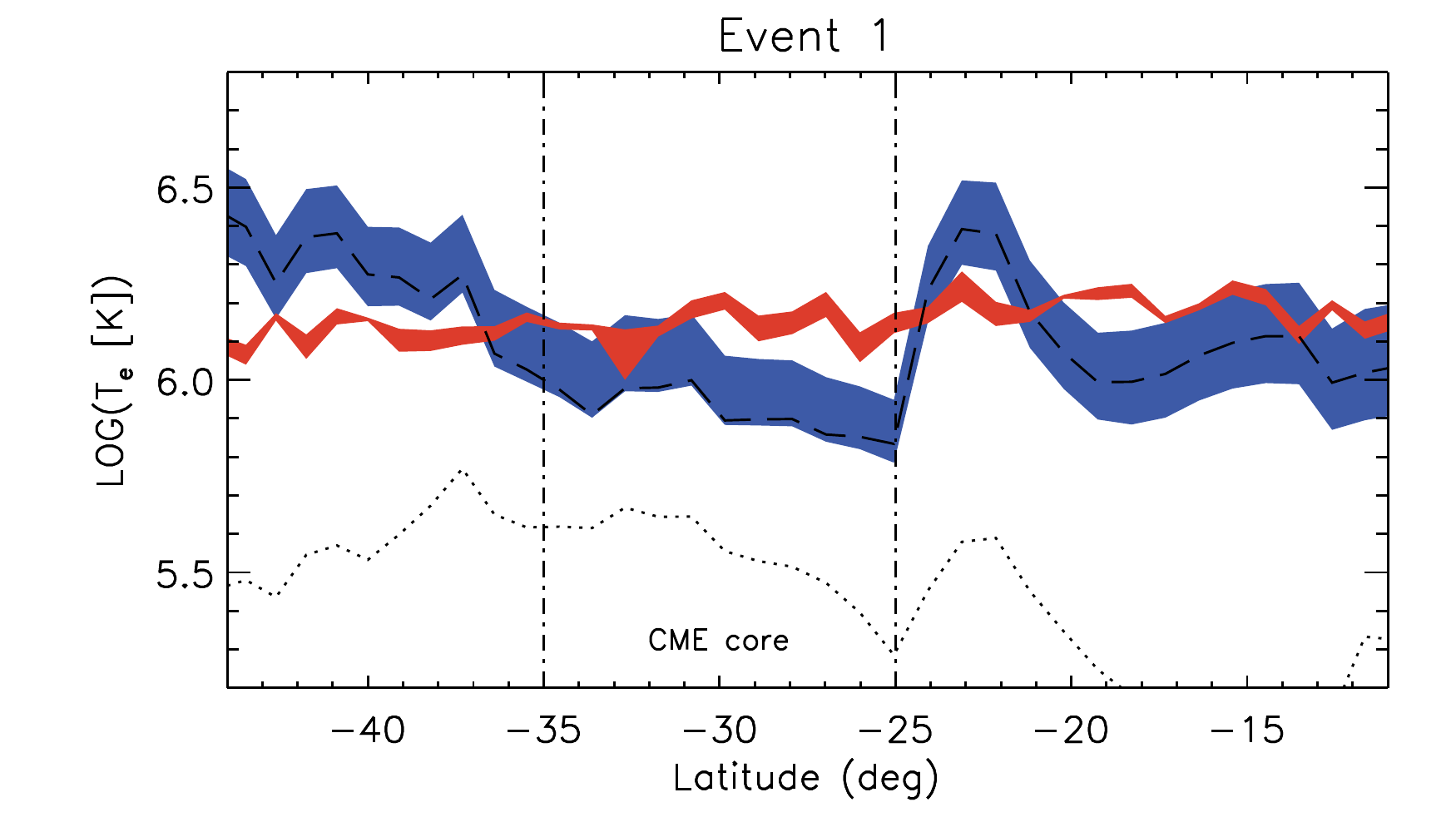}
\includegraphics[width=0.5\columnwidth]{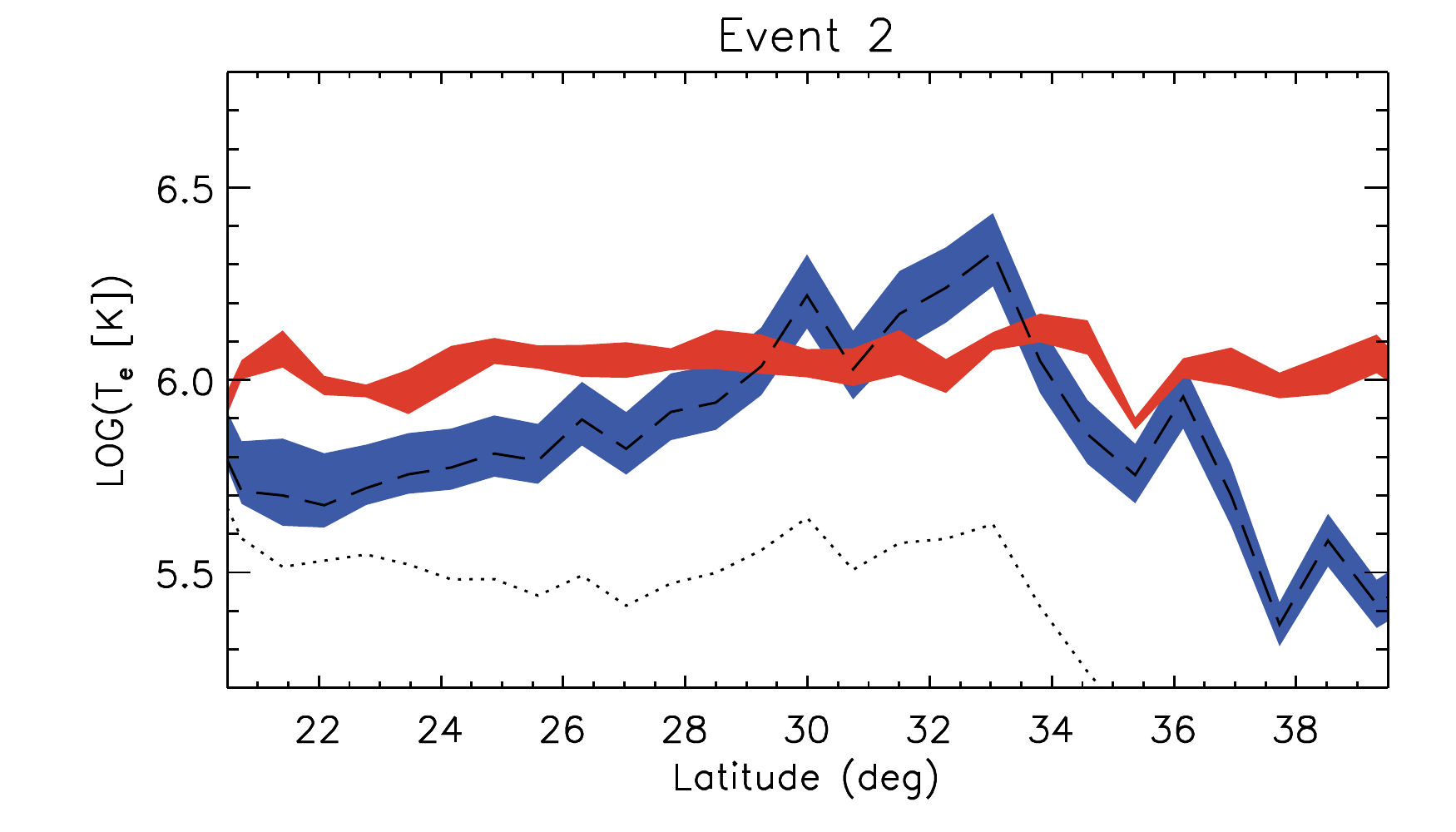}
\includegraphics[width=0.5\columnwidth]{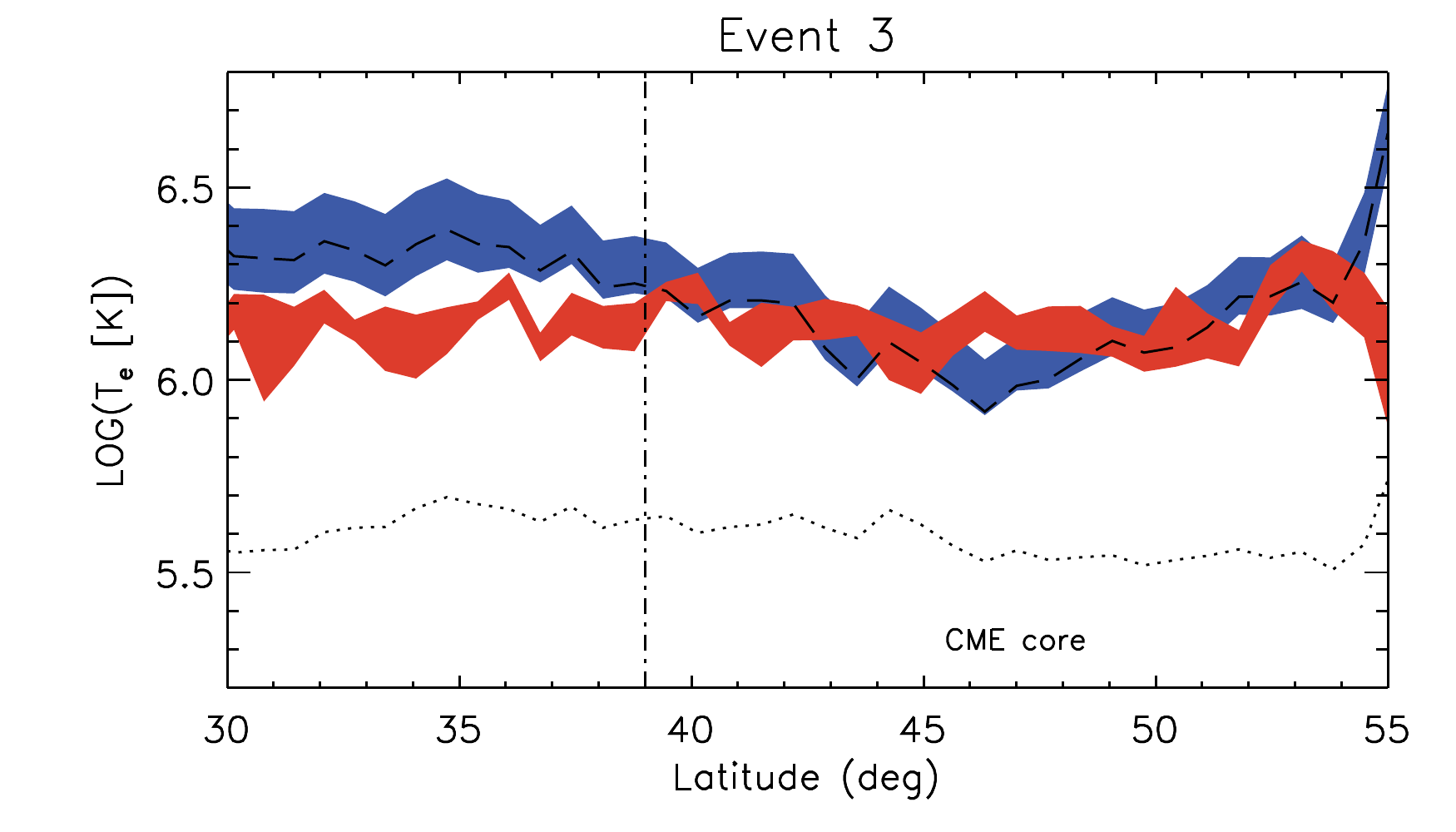}
\end{figure}

The electron temperature profiles along the UVCS slit for the three events are shown in Figure~\ref{temps} (blue shaded areas); these temperatures have been computed from the \lya\ intensity, with the same technique that will be applied to future Metis images. For comparison, these electron temperatures $T_e$ are compared with the effective temperatures $T_{\rm eff}$ derived from the analysis of the UVCS \lya\ spectra as explained in Section~\ref{uvcs_data} (red areas). For both temperatures the width of the shaded areas is equal to the uncertainties affecting the quantities. As mentioned, $T_{\rm eff}$ values should be considered as an upper limit to real hydrogen kinetic temperatures, which is representative, in turn, of the proton temperature $T_p$.
For reference purposes, we also plotted the electron temperature curves derived assuming that the observed \lya\ intensity was due to the radiative component alone (the ``radiative'' approximation; dashed line) and to the collisional component alone (the ``collisional'' approximation; dotted line).
Finally, all the plotted quantities were obtained assuming $L=0.25$~\rsun.

The \lya\ effective temperatures $T_{\rm eff}$ are in all cases quite uniform, within the uncertainties, in the considered regions along the UVCS FOV. 
The average value $T_{\rm eff}\simeq10^6$~K is characteristic of coronal conditions.
The \hi\ \lya\ profile in CME cores is usually narrower than typical coronal profiles and UVCS observations often imply proton temperatures of the order of $10^5$~K \citep{koh06}, well below the value derived in this work. These lower temperatures are usually interpreted as a signature of chromospheric plasma embedded in the expanding CME core. However, \citet{bem07} also found evidence of hydrogen kinetic temperature around $1.6\times 10^6$~K in their analysis of a CME observed by UVCS on January 31, 2000.

\begin{figure}
\caption{\label{pb_temps} Polar maps (as functions of heliographic latitude and heliocentric distance) of the polarized visible light measured by LASCO-C2	for Event~1 (top row), Event~2 (middle row), and Event~3 (bottom row), processed with the NRGF (left column) and with a filter that highlights emission gradients (right column). The observed \hi\ \lya\ intensities and the derived electron temperatures along the UCVS slit are superimposed on the left and right plots, respectively, using color gradients according to the color bars.}
\centering
\includegraphics[width=0.75\columnwidth]{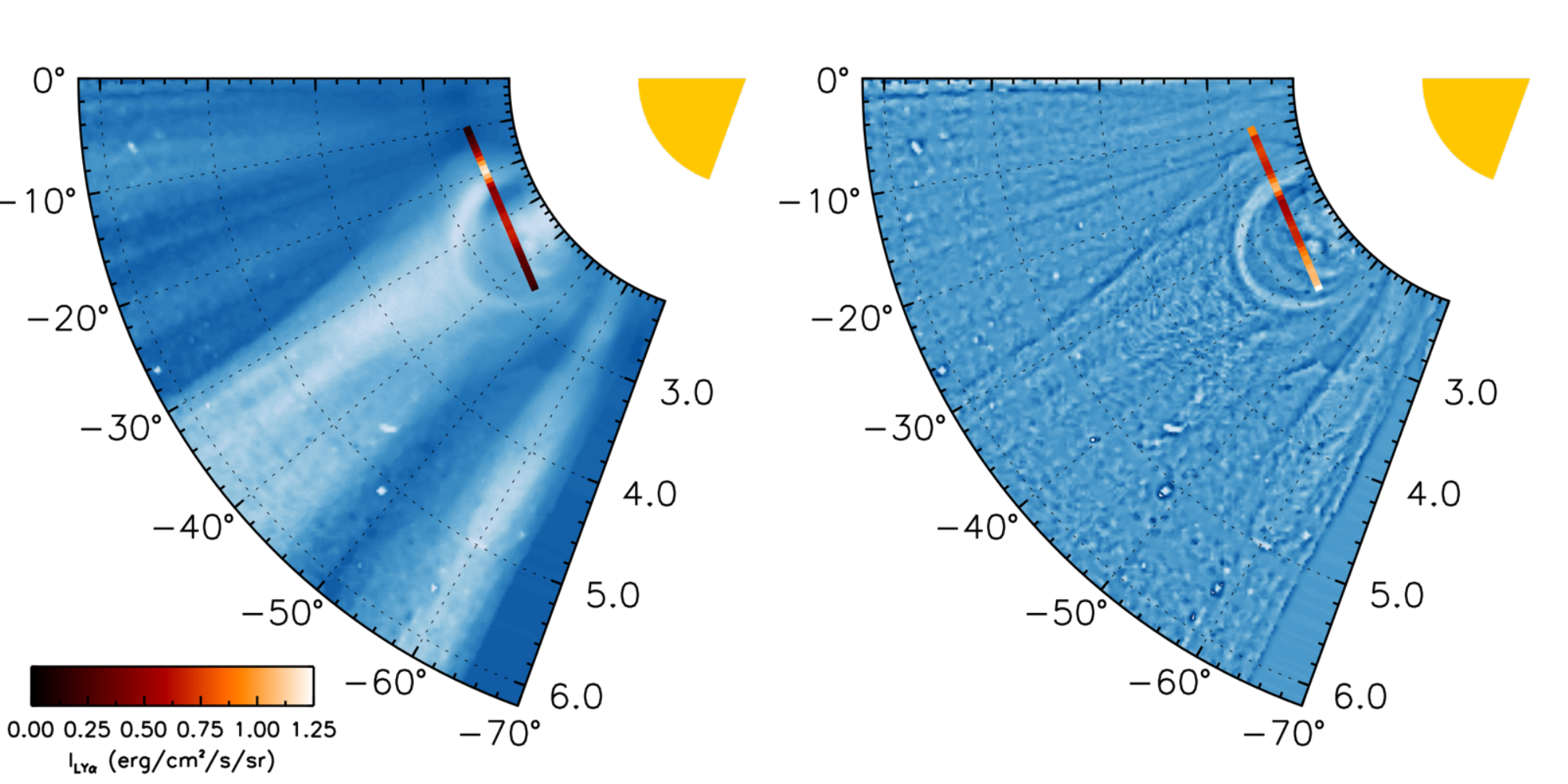}
\includegraphics[width=0.75\columnwidth]{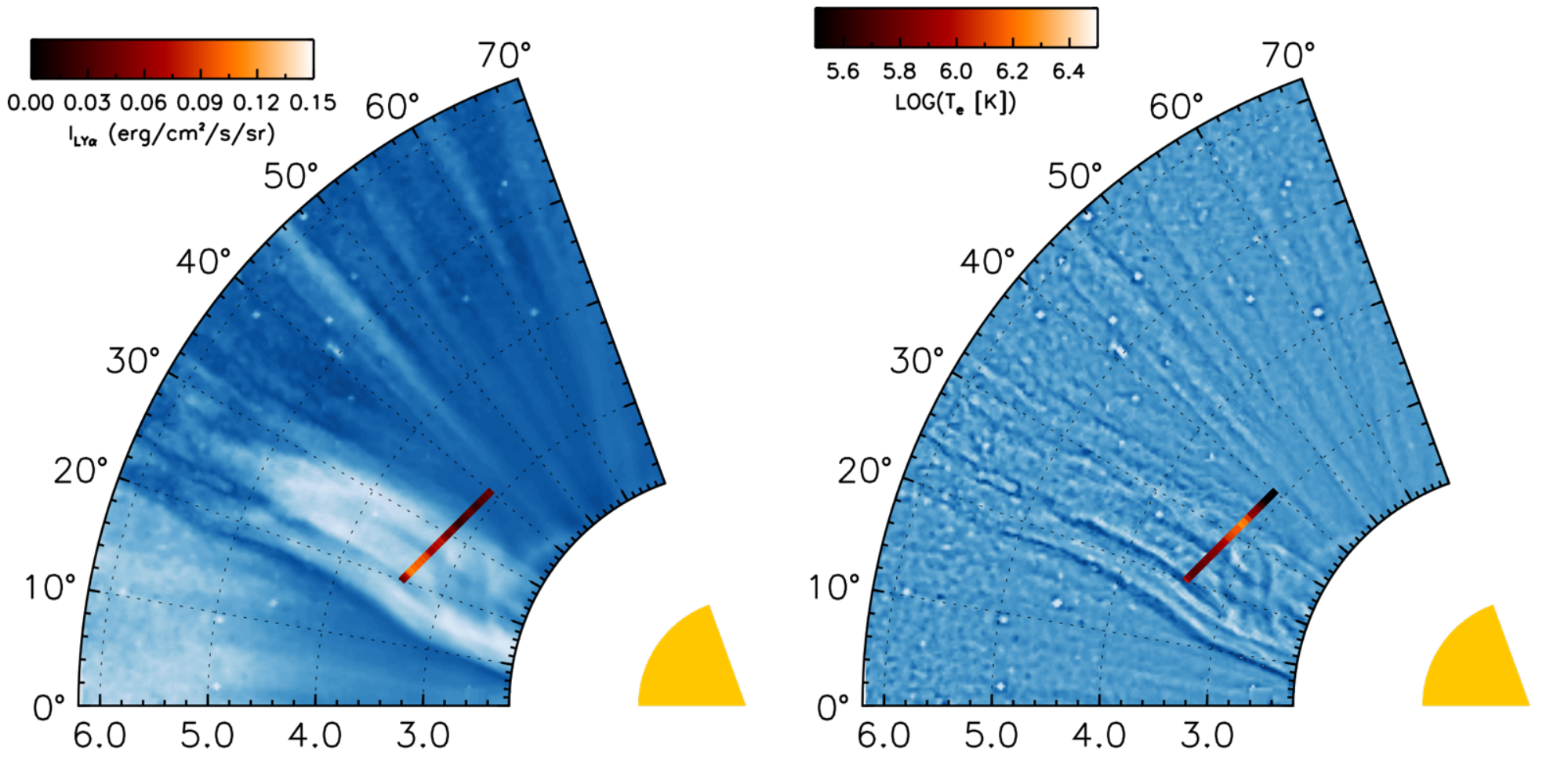}
\includegraphics[width=0.75\columnwidth]{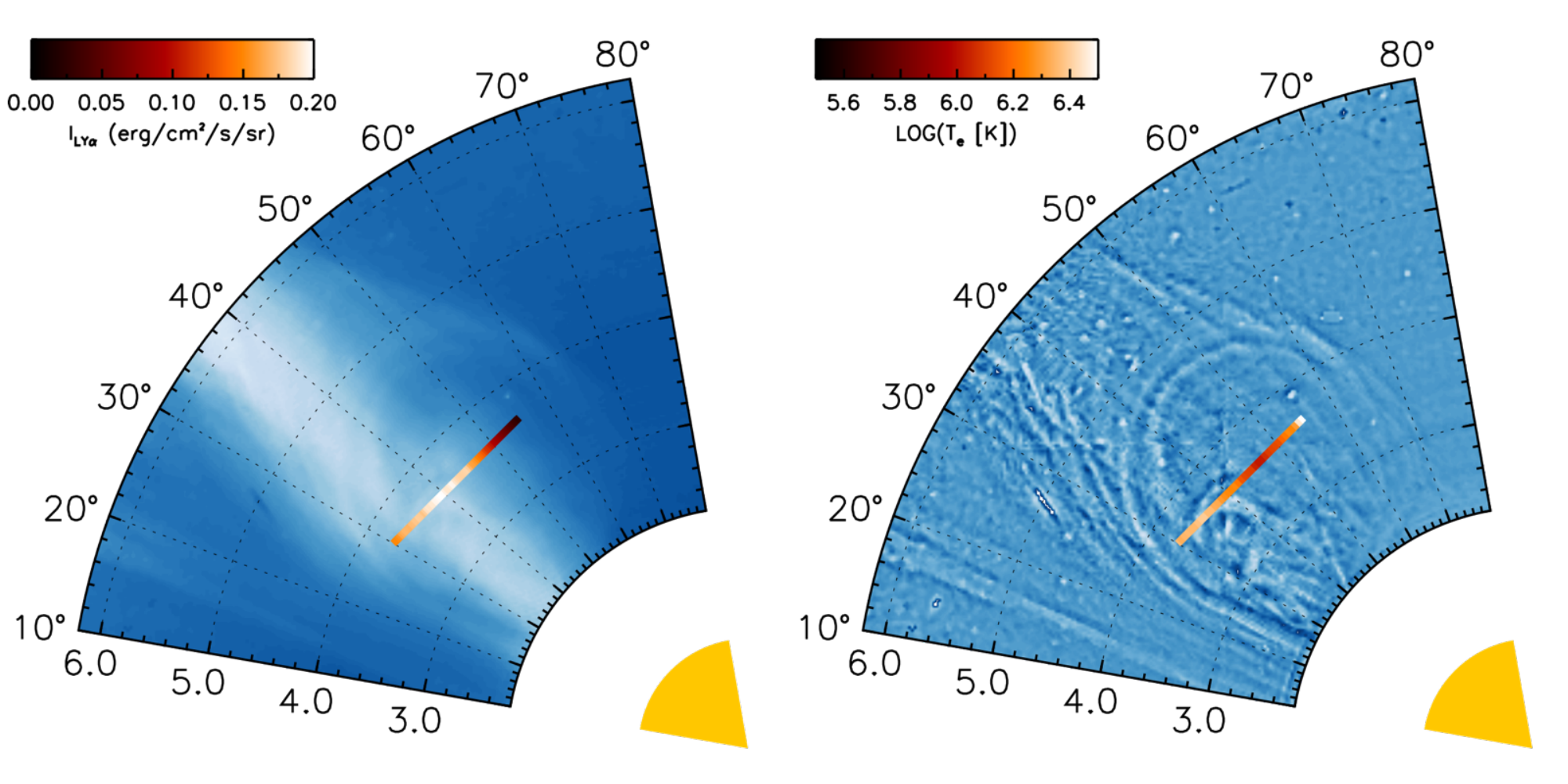}
\end{figure}

The electron temperatures derived in the general case (the blue curves) are in the range $10^{5.5}-10^{6.5}$~K, in agreement with the values typically detected for CMEs.
For Event~1 and~3, where it is possible to identify almost clearly the front and the core of the CME along the UVCS slit based on the distribution of both \lya\ intensity and electron density (see also Fig.~\ref{densities}), the latitudinal variations in the $T_e$ curve reflect to some extent the CME structures.
To make evident possible correlations, we plotted in Figure~\ref{pb_temps} a comparison between the observed \lya\ intensities along the UCVS slit superimposed to the LASCO-C2 $pB$ intensity images (left column), and a comparison between the derived electron temperatures along the UVCS slit and the same LASCO-C2 $pB$ intensity images filtered to highlight emission gradients associated with the CME features (right column).

The comparison shows that in the first case (Event~1, first row), the electron temperature is lower ($\sim 10^6$~K) in the region corresponding to the core, then it increases up to $10^{6.5}$~K in the surrounding regions located in the CME front.
The core $T_e$ is quite large compared to typical temperatures measured by UVCS in the cores of other CMEs \citep[$10^{4.5}-10^{5.5}$~K; see][]{cia00,cia01,akm01,koh06}, but it is similar to the temperature measured, for instance, by \citet{cia03} in the core of a CME observed in 2000.
The peak located around $-22.5^\circ$S is consistent with the position of the CME front and could be interpreted as the signature of heating by plasma compression.
The temperature increase below $-35^\circ$S could be associated with the CME void that is crossed by the UVCS slit: for instance, \citet{cia03} have found evidence of $T_e>10^{6.2}$~K in the void of the same CME mentioned above.

The same correlation can be recognized also in the results of Event~3, although the variations of the electron temperature with latitude are less pronounced in this case.
At latitudes around $30^\circ$N the UVCS slit crosses the front of the CME (see Fig.~\ref{pb_temps}, bottom panels) and correspondingly the electron temperature is higher at those latitudes ($T_e\simeq 10^{6.3}$~K); then it decreases in the regions identifiable as the core of the CME (above $40^\circ$N, where $T_e\approx10^6$~K).
Even in this case we get temperatures quite higher than typical values reported in the literature.

Event~2 is quite peculiar because it does not consist in a classical CME, as described in Section~\ref{events}, thus it is not possible to identify a core or a front region.
However, the steep increase of the electron temperature from $\sim 10^{5.7}$~K at $22^\circ$N to $10^{6.3}$~K at $33^\circ$N is an indication of the different plasma conditions sampled by the UVCS slit.
The southern edge of the slit is superimposed to the ejected blob, as it is evident from Figure~\ref{pb_temps} (middle panels), which most probably consists of plasma at chromospheric temperatures ($\sim 10^5$~K) coming from lower layers of the Sun's atmosphere; as the latitude increases, the slit progressively samples the quiet regions surrounding the blob, which likely have higher temperatures, closer to typical coronal values of $\sim 10^6$~K.
Note that the sudden decrease of $T_e$ above $\sim 33^\circ$N is due to an analogous decrease of the electron density (see Fig.~\ref{densities}) caused by background-subtraction effects, and it could not necessarily imply a real trend.

Comparison with the temperature curves derived with the radiative and the collisional approximations (Figure~\ref{temps}, dashed and dotted lines respectively) shows that in all the three cases the contribution of the radiative component to the observed \lya\ intensity is overwhelming.
This is not surprising, considering that at the quite low velocities characteristic of our events ($v_{\rm out}<250$~\kms\ in all cases) a non negligible fraction ($\sim 20\%$) of the radiative component survives the Doppler dimming effect.
The electron temperatures derived with the collisional approximation, conversely, are very low, around $\sim 10^{5.5}$~K at most.
They are closer to the $T_e$ derived in the general case in the cores of Events~1 and~3 and in the center of the ejected blob of Event~2, but in any case they are a factor of $\sim 2-3$ lower.
These low values could be the result of an under-estimate of the electron density: higher densities would in fact produce a greater collisional component (see Fig.~\ref{model}) and, in turn, higher temperatures would be necessary to reproduce the same observed intensity.
However, in our analysis the electron density is constrained by the column density derived from the visible light (through Eq.~2); to obtain greater values of $\avg{n_e}$ we should reduce the value of $L$; nevertheless, an extension of the CME structures along the LOS lower than 0.25~\rsun\ is quite  unrealistic (this value is already 4 times lower than the typical assumption of $L=1$~\rsun\ usually done in similar calculations).
Note that, since the radiative component depends directly on the electron column density, the choice on the value of $L$ does not affect substantially our final results in the general case.

Finally, as far as the effective temperatures derived from UVCS \lya\ spectra can be considered as an indication of the \hi\ kinetic temperatures, our results suggest that electron temperatures are quite comparable with hydrogen temperatures, in particular in the regions where the former are lower.

\section{Discussion and conclusions}

In this work we demonstrate how future observations of CMEs that will be provided by Metis coronagraph on-board Solar Orbiter will be analyzed to infer physical parameters of the plasma involved in the eruption and to derive, in particular, the electron temperature of the ejected gas. 
To this end, we first selected eruptive events observed at the same times and at the same coronal locations both in LASCO-C2 polarized-brightness images, and in the UV \hi\ \lya\ line by UVCS; it turns out that over the whole era of UVCS, only three events were sampled by both instruments. 
Then, we analyzed VL data and showed that they can be used to derive not only the electron column density and volumetric density inside the CME, but also the average location of the emitting plasma along the LOS. 
This in turn can be used, on one hand to better constrain the plasma electron densities, on the other hand, to measure the CME propagation direction with respect to the POS and to derive the unprojected CME speed at different latitudes (i.e., along the UCVS slit).

The unprojected speeds can be converted into Doppler-dimming factors, the missing ingredient needed to combine VL with UV intensities. 
Given the electron densities and Doppler-dimming factors, the combination of VL and UV intensities provides an estimate of CME plasma electron temperatures. 
As we showed here, for the three events we selected the unprojected speeds were so small that the \lya\ emission is still dominated by the radiative component. Nevertheless, we expect that for major and faster CMEs the situation could be even reversed, with the \lya\ emission being dominated by the collisional component, in particular in the denser parts such as the CME cores. 
In this work we found that the CME cores are usually associated with cooler plasma, and that a significant rise of temperatures is observed moving from the core to the front of the CME.

The determination of electron temperatures inside CMEs is of crucial importance. 
In fact, one of the main problems left open after the UVCS era is the real evolution of the CME thermal energy during their expansion. 
Different authors found that during the expansion additional heating sources need to be considered in order to reproduce the observed UV emission, with heating rates comparable \citep{akm01, murphy2011} or even larger \citep{lee2009, landi2010} than kinetic and potential energies carried by the CME. 
Furthermore, \citet{bem07} demonstrated with the only existing multi-slit study of a CME based on UVCS data, that the CME plasma temperature is increasing during the expansion, implying again the existence of an additional thermal-energy source. 
Same results have been recently confirmed by \citet{lee2015} using Hinode/XRT images. 
Nevertheless, a clear interpretation for the source of this additional thermal energy is missing so far.

In this work we demonstrate how CME electron temperatures can be derived using VL images and UV \lya\ intensities. 
Nevertheless, being limited here to the one dimensional FOV of the UVCS slit and to the single time when both VL $pB$ and UV \lya\ emissions were observed, it was not possible to study neither the CME plasma temperature distribution within the whole CME bubble, nor its time evolution during the CME expansion. 
Full investigation of these aspects will be possible thanks to future data that will be provided by the Metis coronagraph on board the Solar Orbiter. 
In fact, sequences of VL and UV images that will be acquired at the same time, will allow to study both the thermal energy distribution within the CME bubble at a given time, and its evolution during the CME propagation. 
Further analysis on the kinematics and 3D structure of CMEs will be possible also thanks to the synergies between Metis and other next-generation coronagraphs, such as ASPIICS \citep[][]{renotte15} on board the ESA Proba-3 mission.
Moreover, for events that will be observed during quadratures also by other spacecrafts (such as Solar Probe Plus), the combination of these information with \textit{in situ} measurements made close to the Sun will allow to tightly constrain the temporal evolution of thermal energy of the ejected plasma during its early interplanetary propagation, thus letting connections with the still open issue of \textit{in situ} detections of high-ionization states of heavy ions in interplanetary CMEs.

\begin{acknowledgements}
The work of RS has been funded by the Agenzia Spaziale Italiana through contracts ASI/INAF N. I/013/12/0 and I/013/12/0-1.
Mauna Loa Mark-{\sc iv} data are courtesy of the Mauna Loa Solar Observatory, operated by the High Altitude Observatory, as part of the National Center for Atmospheric Research (NCAR). NCAR is supported by the National Science Foundation. {\it SOHO} is a mission of international cooperation between ESA and NASA. 
\end{acknowledgements}

\clearpage

\clearpage


\begin{thebibliography}{}
\bibitem[Akasofu(2011)]{akasofu_2011} Akasofu, S.-I. 2011, Sp. Sci. Rev., 164, 85
\bibitem[Akmal et al.(2001)]{akm01} Akmal, A., Raymond, J. C., Vourlidas, A., et al. 2001, \apj, 553, 922
\bibitem[Antonucci et al.(2012)]{antonucci12} Antonucci, E., Fineschi, S., Naletto, G., et al. 2012, Proc. of the SPIE, 8443, id. 844309
\bibitem[Airapetian \& Usmanov(2016)]{airapetian2016} Airapetian, V. S. \& Usmanov, A. V. 2016, \apjl, 817, id. L24
\bibitem[Bemporad et al.(2007)]{bem07} Bemporad, A., Raymond, J. C., Poletto, G., \& Romoli, M. 2007, \apj, 655, 576
\bibitem[Bemporad et al.(2010)]{bemporad2010} Bemporad, A., Soenen, A., Jacobs, C., et al. 2010, \apj, 718, 251
\bibitem[Bemporad et al.(2011)]{bemporad2011} Bemporad, A., Mierla, M., \& Tripathi, D. 2011, A\&A, 531, id.A147
\bibitem[Bemporad et al.(2014)]{bemporad14} Bemporad, A., Susino, R., \& Lapenta, G. 2014, \apj, 784, id. 102
\bibitem[Bemporad \& Pagano(2015)]{bem15} Bemporad, A. \& Pagano, P. 2015, A\&A, 576, A93
\bibitem[Billings(1966)]{bil66} Billings, D. E. 1966, A guide to the Solar Corona (New York: Academic Press)
\bibitem[Brueckner et al.(1995)]{brueckner95} Brueckner, G. E., Howard, R. A., Koomen, M. J., et al. 1995, Sol. Phys., 162, 357
\bibitem[Byrne et al.(2010)]{byrne2010} Byrne, J. P., Maloney, S. A., McAteer, R. T. J., et al. 2010, Nature Comm., v. 1, id. 74
\bibitem[Ciaravella et al.(2000)]{cia00} Ciaravella, A., Raymond, J. C., Thompson, B. J., et al. 2000, \apj, 529, 575
\bibitem[Ciaravella et al.(2001)]{cia01} Ciaravella, A., Raymond, J. C., Reale, F., Strachan, L., \& Peres, G. 2001, ApJ, 557, 351
\bibitem[Ciaravella et al.(2003)]{cia03} Ciaravella, A., Raymond, J. C., van Ballegooijen, A., et al. 2003, \apj, 597, 1118
\bibitem[Ciaravella et al.(2005)]{cia05} Ciaravella, A., Raymond, J. C., Kahler, S. W., et al. 2005, \apj, 621, 1121
\bibitem[Dere et al.(2005)]{dere05} Dere, K. P., Wang, D., \& Howard, R. 2005, \apjl, 620, L119
\bibitem[Dobrzyzca et al.(2003)]{dob03} Dobrzycka, D., Raymond, J. C., Biesecker, D. A., et al. 2003, \apj, 588, 586
\bibitem[Farrugia et al.(2006)]{far06} Farrugia, C. J., Jordanova, V. K., Thomsen, M. F., et al. 2006, JGR (Space Physics), 111, 11104
\bibitem[Fineschi et al.(2012)]{fineschi12} Fineschi, S., Antonucci, E., Naletto, G., et al. 2012, Proc. of the SPIE, 8443, id. 84433H
\bibitem[Frazin et al.(2012)]{frazin12}  Frazin, R. A., V\'asquez, A. M., Thompson, W. T., et al. 2012, Sol. Phys. 280, 273
\bibitem[Gouttebroze et al.(1978)]{gou78} Gouttebroze, P., Lemaire, P., Vial, J.-C., \& Artzner, G. 1978, ApJ, 225, 655
\bibitem[Gurnett et al.(2015)]{gurnett2015} Gurnett, D. A., Kurth, W. S., Stone, E. C., et al. 2015, \apj, 809, id. 121
\bibitem[Heinzel et al.(2016)]{heinzel16} Heinzel, P.., Susino, R.., Jej\v{c}i\v{c}, S., et al. 2016, Astron. \& Astrophys., 589, id. A128
\bibitem[Hultqwist(2008)]{hultqwist_2008} Hultqwist, B. 2008, Journ. of Atmosp. and Sol.-Terr. Phys., 70, 2235
\bibitem[Isavnin et al.(2014)]{isavnin2014} Isavnin, A., Vourlidas, A., \& Kilpua, E. K. J. 2014, Sol. Phys., 289, 2141
\bibitem[Iju et al.(2014)]{iju2014} Iju, T., Tokumaru, M., \& Fujiki, K. 2014, Sol. Phys., 289, 2157
\bibitem[Kay et al.(2013)]{kay2013} Kay, C., Opher, M. \& Evans, R. M. 2013, \apj, 775, 5
\bibitem[Kohl \& Withbroe(1982)]{koh82} Kohl, J. L. \& Withbroe, G. L. 1982, ApJ, 256, 263
\bibitem[Kohl et al.(1995)]{kohl1995} Kohl, J. L., Esser, R., Gardner, L. D., et al. 1995, Sol. Phys., 162, 313
\bibitem[Kohl et al.(2006)]{koh06} Kohl, J. L., Noci, G., Cranmer, S. R., \& Raymond, J. C. 2006, Astron. Astrophys. Rev., 13, 31
\bibitem[Landi et al.(2010)]{landi2010} Landi, E., Raymond, J. C., Miralles, M. P., \& Hara, H. 2010, \apj, 711, 75
\bibitem[Landi Degl'Innocenti \& Landolfi(2004)]{lan04} Landi Degl'Innocenti, E., \& Landolfi, M. 2004, Polarization in Spectral Lines (Dordrecht: Kluwer Academic Publishers)
\bibitem[Lee et al.(2009)]{lee2009} Lee, J.-Y., Raymond, J. C., Ko, Y.-K., \& Kim, K.-S. 2009, \apj, 692, 1271
\bibitem[Lee et al.(2015)]{lee2015} Lee, J.-Y., Raymond, J. C., Reeves, K. K., et al. 2015, \apj, 798, id. 106
\bibitem[Lemaire et al.(2002)]{lem02} Lemaire, P., Emerich, C., Vial, J.-C., et al. 2002, ESA SP-508, 219
\bibitem[Llebaria \& Lamy(2008)]{llebaria08} Llebaria, A. \& Lamy, P. 2008, Proc. of the SPIE, 7010, id. 70101I 
\bibitem[Lockwood(1971)]{lockwood1071} Lockwood, J. A. 1971, Sp. Sci. Rev., 12, 658, 1971
\bibitem[Mierla et al.(2009)]{mie09} Mierla, M., Inhester, B., Marqu\'e, C., et al. 2009, Sol. Phys., 259, 123
\bibitem[Moran \& Davila(2004)]{mor04} Moran, T. G., \& Davila, J. M. 2004, Science, 305, 66
\bibitem[Moran et al.(2006)]{mor06} Moran, T. G., Davila, J. M., Morrill, J. S., Wang, D., \& Howard, R. 2006, So. Phys., 237, 211
\bibitem[M\"{o}stl et al.(2015)]{mostl2015} M\"{o}stl, C., Rollett, T., Frahm, R. A., et al. 2015, Nature Comm., v. 6, id. 7135
\bibitem[Murphy et al.(2011)]{murphy2011} Murphy, N. A., Raymond, J. C., \& Korreck, K. E. 2011, \apj, 735, id. 17
\bibitem[Noci et al.(1987)]{noc87} Noci, G., Kohl, J. L., \& Withbroe, G. L. 1987, ApJ, 315, 706
\bibitem[Noci et al.(1993)]{noc93} Noci, G., Poletto, G., Suess, S. T., et al. 1993, Sol. Phys., 147, 73
\bibitem[Pagano \& Bemporad(2015)]{pag15} Pagano, P., Bemporad, A., \& Mackay, D. H.  2015, ApJ, 582, A72
\bibitem[Qu\'emerais \& Lamy(2002)]{quemerais02} Qu\'emerais, E. \& Lamy, P. 2002, A\&A, 393, 295
\bibitem[Raymond \& Ciaravella(2004)]{ray04} Raymond, J. C. \& Ciaravella, A. 2004, \apjl, 606, L159 
\bibitem[Renotte et al.(2015)]{renotte15} Renotte, E., Alia, A., Bemporad, A., et al. 2015, Proc. of the SPIE, 9604, id. 96040A
\bibitem[Romoli et al.(2014)] {romoli14} Romoli, M., Landini, F., Antonucci, E., et al. 2014, Proc. of the ICSO 2014 Conference, Tenerife (Spain), 7-10 October 2014
\bibitem[Susino et al.(2014)]{susino14} Susino, R., Bemporad, A., \& Dolei, S. 2014, \apj, 790, id. 25
\bibitem[Susino et al.(2015)]{susino15} Susino, R., Bemporad, A., \& Mancuso, S. 2015, \apj, 812, id. 119
\bibitem[Temmer et al.(2011)]{temmer2011} Temmer, M., Rollett, T., M\"{o}stl, C., et al. 2011, \apj, 743, id. 101
\bibitem[Thompson et al.(2012)]{thompson2012} Thompson, W. T., Kliem, B., \& T\"{o}r\"{o}k, T. 2012, Sol. Phys., 276, 241
\bibitem[Tobiska et al.(1997)]{tob97} Tobiska, W. K., Pryor, W. R., \& Ajello, J. M. 1997, Geo. Res. Lett., 24, 1123 
\bibitem[Vernazza \& Reeves(1978)]{ver78} Vernazza J. E. \& Reeves, E. M. 1978, \apjs, 37, 485
\bibitem[Vourlidas et al.(2000)]{vou00} Vourlidas, A., Subramanian, P., Dere, K. P., \& Howard, R. A. 2000, ApJ, 534, 456
\bibitem[Withbroe et al.(1982)]{wit82} Withbroe, G. L., Kohl, J. L., Weiser, H., \& Munro, R. H. 1982, Space. Sci. Rev., 33, 17
\bibitem[Woods et al.(2000)]{woo00} Woods, T. N., Tobiska, W. K., Rottam, G. J., \& Worden, J. R. 2000, JGR, 105, 27217
\bibitem[Wu et al.(2007)]{wu07} Wu, C.-C., Fry, C. D., Dryer, M., et al. 2007, Adv. Sp. Res., 40, 1827
\end{thebibliography}
\end{document}